%% file: Monadic_Dynamics.tex
\documentclass[11pt]{article}
\input{modules/packages}

\input{modules/macros}

\begin{document}

%% TITLE %%	
	\title{\textbf{Monadic dynamics}}
	\author{Stefano Gogioso\\
		\footnotesize{Quantum Group}\\ 
		\footnotesize{Department of Computer Science}\\
		\footnotesize{University of Oxford, UK}\\
		\href{mailto:stefano.gogioso@cs.ox.ac.uk}{\small{stefano.gogioso@cs.ox.ac.uk}}
	}
	%\date{}
	\maketitle
	
%% ABSTRACT AND TABLE OF CONTENTS %%
	\begin{abstract}
	
		We develop a monadic framework formalising an operational notion of dynamics, seen as the setting and evolution of initial value problems, in general physical theories. We identify in the Eilenberg-Moore category the natural environment for dynamical systems and characterise Cauchy surfaces abstractly as automorphisms in the Kleisli category.\\

		Our main results formally vindicates the Aristotelian view that time and change are defined by one another. We show that dynamics which respect the compositional structure of physical systems always define a canonical notion of time, and give the conditions under which they can be faithfully seen as actions of time on physical systems.\\

		Finally, we construct state spaces and path spaces, and show that our framework to be equivalent to the path space approaches to dynamics. The monadic standpoint is thus as strong as the established paradigms, but the shift from histories to dynamics helps shed new light on the nature of time in physics.\\

		In the appendix we present some additional structures of wide applicability, introduce propagators and draft applications to quantum theory, classical mechanics and network theory.

	\end{abstract}
	\newpage

	\tableofcontents
	\newpage

	\setlength{\parindent}{0pt}
	\numberwithin{equation}{section}

%% BEGIN SECTION - Introduction %%
\section{Introduction}
	\label{section_GeneralPhilosophy}

	The operational understanding of time as some sort of \textit{universal}, or \textit{free}, notion of change has influenced the philosophy of science ever since Aristotle wrote it into history with his \textit{Physics}: when talking about the nature of time in book IV (ch.12, 220b 14-15), the Greek father of empiricism indeed claims that \inlineQuote{\textit{not only do we measure change by time, but time by change, because they are defined by one another}}. We decide to take the operational notion of time as free dynamics seriously in this work, employing monads as a uniform language for the treatment of time and dynamics, seen as the setting and evolution of initial value problems, in general physical theories.\\

	In section \ref{section_GeneralFramework}, the \textit{monadic dynamics} framework will be progressively exemplified in the mathematical theory of dynamical systems, which has a simple and familiar structure in which to showcase the main ideas. The framework is however much more general, and what will be referred to as \textit{time} in this work actually encompasses a breadth of different notions of dynamics on closed systems, including all internal groups and monoids acting in symmetric monoidal categories. Furthermore, we take an empirical standpoint and restrict ourselves to notions of time that can be internalised, or simulated, by the physical theory, making our time objects a generalisation of physical clocks.\\

	In section \ref{section_TimeSymmetricMonoidal} we will prove our main claim: under some physically sensible conditions, time is defined by dynamics and dynamics can always be seen as actions of time on physical systems. In section \ref{section_SpaceConcreteHistories} we will show our approach to be equivalent to the more common path-space approaches. The main open questions then becomes: what operational characterisation of the dynamics will force time to take the \inlineQuote{linear} form observed in most theories of physics? Only time will tell.
%% END SECTION - Introduction %%

%% BEGIN SECTION - An operational notion of time %%
\section{An operational notion of time}
	\label{section_GeneralFramework}

	\subsection{A categorical approach to physics}
		\label{section_SpaceFreeHistories}
		\label{section_CategoricalApproach}

		Under the Wittgensteinian \footnote{In fact seemingly traceable to Gilbert Ryle's \textit{The theory of meaning}, rather than to Wittgenstein himself.} slogan \inlineQuote{\textit{Don't ask for the meaning, ask for the use}}, the categorical approach to physics shifts the attention from the physical systems with their structure to the ways the systems relate to and transform into each other. The physical systems themselves become mere labels, their internals nothing other than an emergent characterisation of the way they transform.\footnote{From a mathematical point of view, we see morphisms as \textit{structure-inducing} rather than \textit{structure-preserving}, the latter notion being better represented by functoriality and naturality.}\\

		Once we have a category of physical systems representing a specific theory, it is natural to ask if and how the dynamics of those systems can be described within the category, i.e. with the language and structure of the theory itself. We assume that the concrete dynamics of a physical system $A$ share enough common structure (i.e. a notion of \inlineQuote{time}), which can be abstracted and simulated by some bigger physical system $\monad A$ of our theory, its \textbf{space of free histories}. Each individual dynamic then is a particular way of making the free histories concrete by folding $\monad A$ onto $A$, i.e. turning each free history into a concrete history in a way that is meaningful for the physical theory under consideration:
		\begin{equation}
			\text{dynamic} : \monad A \epim A
		\end{equation}

		In the mathematical theory of dynamical systems, for example, we have a fixed notion of time that works for all systems, and a system $\timeobj = \reals, \naturals, \integers, ...$ that simulates it. The construction of the space of free histories for $A$ will then amount to taking $\monad A = A \cartesianTensor \timeobj$, i.e. the space of \inlineQuote{formal time evolutions} of points / states / elements of $A$. The dynamics of the system will be morphisms $(a,t) \mapsto \evolve{t}{a}$ evolving the system in some consistent manner, which will be elucidated in the rest of this section.

	\subsection{The operational approach to dynamics}
		\label{section_OperationalApproach}

		We operationally characterise time by seeing it as a way of changing, evolving a system: to do so, we consider the original system $A$, its space of free histories $\monad A$ and the space $\monad \monad A$ of free histories of $\monad A$ (as the latter is itself a physical system).

		\subsubsection{Free and concrete histories}
		\label{section_freeConcreteHistoriesGeneral}
		Under the structure-inducing view of the category $\CategoryC$ of physical systems for a theory, when talking about subsystems of a system $A$ we will mean ways $D \stackrel{d}{\rightarrow} A$ of transforming other systems to it. The ways of transforming a fixed system $D$ to system $A$ form the space $\Hom{\CategoryC}{D}{A}$ of \textbf{$D$-points} of $A$, and the \textbf{sheaf of subsystems} of a physical system $A$ will be: 
		\begin{equation}
			\Subsys{}{A} \eqdef \Hom{\CategoryC}{\emptyArg}{A}
		\end{equation}
		To us, \textbf{subsystems} of $A$ will be general morphisms $D \rightarrow A$: we will avoid the nomenclature \inlineQuote{subobjects} altogether, and refer to monomorphisms $D \monom A$ as \textbf{faithful subsystems}. The latter are subsystems where all the structure of $D$ is faithfully embedded into structure for $A$. Please note that, from now on, we will confuse the hom functors $\Hom{\CategoryC}{\emptyArg}{A}$ and $\Hom{\CategoryC}{A}{\emptyArg}$, their categories of elements, and the slice and coslice categories $\CategoryC / A$ and $A \backslash \CategoryC$. \\ 

		We define the \textbf{free history} of some subsystem $D \stackrel{d}{\rightarrow} A$ to be the following subsystem of $\monad A$:
		\begin{equation}
			\freehistory{d} \eqdef \monad d: \monad D \rightarrow \monad A
		\end{equation}
		The \textbf{free histories} contribute, as expected, to the structure of the physical system $\monad A$:
		\begin{equation}
			\FreeHistories{}{A} \eqdef \suchthat{\freehistory{d}}{d : \Subsys{}{A}} \inject \Subsys{}{\monad A}
		\end{equation}

		\subsubsection{Concrete histories}
		\label{section_ConcreteHistoriesDef}
		Each \textbf{concrete history} of a subsystem $D \stackrel{d}{\rightarrow} A$ will be obtained as the image of its free history under a dynamic $\alpha: \monad A \epim A$ for the parent system $A$:
		\begin{equation}
			\concretehistory{\alpha}{d} \eqdef \alpha \cdot \freehistory{d} : \monad D \rightarrow A
		\end{equation}
		The \textbf{space of concrete histories} $\ConcreteHistories{}{A}$ for a system $A$ will be discussed in section \ref{section_SpaceConcreteHistories}.

		\subsubsection{Liftings}
		When talking about the \textbf{lifting} of a morphism $f: A \rightarrow B$, we mean the following morphism:
		\begin{equation}
			\monad f : \monad A \rightarrow \monad B
		\end{equation}
		In this sense, the free histories for $A$ are the liftings of subsystems of $A$. But when using the work \inlineQuote{lifting}, we will often have in mind its \textbf{push-forward} action $\pushforward{f}$ on the free histories of systems:
		\begin{equation}
			\begin{array}{cccc}
			\pushforward{f} \eqdef \monad f \cdot \emptyArg: & \FreeHistories{}{A} & \rightarrow & \FreeHistories{}{B} \\
			& \freehistory{d} &\mapsto & \freehistory{f \cdot d}
			\end{array}
		\end{equation}
		There is also a \textbf{pullback} action $\pullback{h}$ of liftings on the free histories of systems:

		\begin{equation}
			\begin{array}{cccc}
			\pullback{h} \eqdef \emptyArg \cdot \monad h: & \FreeHistories{}{A} & \rightarrow & \FreeHistories{}{A} \\
			& \freehistory{d} &\mapsto & \freehistory{d\cdot h}
			\end{array}
		\end{equation}

		\subsubsection{Canonical initial surface}
		If we want to characterise \inlineQuote{time as change}, we first need a starting point to change the system from, i.e. an \textbf{initial surface} $\unit{A}$ embedding $A$ as a faithful subsystem of $\monad A$:
		\begin{equation}
			\unit{A}: A \monom \monad A
		\end{equation}
		For the mathematical theory of dynamical systems, this is given by taking the system at time zero:
		\begin{equation}
		\unit{A} = a \mapsto (a,0)
		\end{equation}

		\subsubsection{Canonical evolution}
		The free histories of $A$ have a \textbf{canonical dynamic / evolution}, in that they encode the abstract notion of free dynamics for $\monad A$. Operationally, we need a way of canonically evolve the free histories of $A$ by making them follow themselves, in the form of a dynamic for $\monad A$:
		\begin{equation}
			\mult{A}: \monad \monad A \epim \monad A
		\end{equation}
		For the mathematical theory of dynamical systems, this is given by adding the times: 
		\begin{equation}
			\mult{A} = ((a,t),\Delta t) \mapsto (a,t+\Delta t)
		\end{equation}

		\subsubsection{Structural rigidity}
		We expect initial surfaces and canonical evolution to be compatible with the structure lifted from $A$ to $\monad A$ and $\monad \monad A$:
		\input{modules/diagrams/GeneralFramework/MonadUnitMultNaturality}
		
		In particular, naturality of $\unit{A}$ means that $A$ can be seen as a leaf in $\monad A$, all its morphisms (and hence its structure) faithfully lifted trough the embedding. For the mathematical theory of dynamical systems, this condition reads: 
		\begin{equation}
			\monad f \, (a,0) = (f(a),0)
		\end{equation}
		This \textbf{leaf-wise action} of the liftings always holds on the initial surface, but in general the space of free histories need not be foliated: the mathematical theory of dynamical systems is, and the liftings act leaf-wise at all times:
		\begin{equation}
			\monad f \, (a,t) = (f(a),t) \text{ for all } t : \timeobj
		\end{equation}

		\subsubsection{Initial value problems}
		\label{section_InitialValueProblems}
		For $\unit{A}$ to act as an initial surface, we have to require that all the concrete dynamics $\alpha: \monad A \epim A$ for the physical system $A$ respect it as such, i.e. do nothing on it:
			\input{modules/diagrams/GeneralFramework/AlgebraInitialSurface}
		This is the abstract construction required to be able to formulate an \textbf{initial value problem}, and for the mathematical theory of dynamical systems it reads:
		\begin{equation}
			\evolve{0}{a_0} = a_0
		\end{equation}

		\subsubsection{Free dynamics}
		\label{section_FreeDynamics}
		The canonical evolution $\mult{A}$ is meant to encode the \textbf{free dynamics} of system $A$: by this we mean that it encodes the abstract compositional aspect of the evolution, i.e. it tells us what it means to \inlineQuote{evolve the system a bit, then evolve it some more}. This can be formalised by the following commutative diagram (more details can be found in section \ref{section_AppendixFreeDynamics} of the appendix):
			\input{modules/diagrams/GeneralFramework/AlgebraFreeDynamicalStructure}
		
		For the mathematical theory of dynamical systems, this condition reads:
		\begin{equation}
			\evolve{\Delta t}{\evolve{t}{a_0}} = \evolve{t+\Delta t}{a_0}
		\end{equation}

	\subsection{Monadic formulation}
		\label{section_ConcreteDynamics}

		\subsubsection{Dynamics as a monad}
		\label{section_ConcreteDynamicsMonad}
		As the reader familiar with the theory of monads will have already noticed, the operational approach presented in section \ref{section_OperationalApproach} (and section \ref{section_AppendixOperationalApproach} of the appendix) is equivalent to asking that $(\monad, \unit{}, \mult{})$ is a monad for the category $\CategoryC$ of physical systems, with dynamics being the algebras of the monad.\\

		Without loss of generality, we will assume each systems to have at least one dynamic: from a formal point of view, this implies that the unit $\unit{A}$ is always (split) mono, and hence the initial surface is a faithful subsystem of the space of free histories.

		\subsubsection{Eilenberg-Moore category and dynamical systems}
		\label{section_ConcreteDynamicsEMCategory}
		The algebras for the monad form, as usual, the \textbf{Eilenberg-Moore category} $\CategoryC^{\monad}$, with objects the bundles $\alpha: \monad A \epim A $ given by the dynamics, and morphisms $f: \alpha \rightarrow \beta$ the bundle morphisms obtained by lifting:
			\input{modules/diagrams/GeneralFramework/EMCategoryMorphisms}
		From now on we will refer to the objects of the Eilenberg-Moore category as \textbf{dynamical systems}; we'll also refer to the morphisms between them as \textbf{morphisms of dynamical systems}, or \textbf{dynamical transformations}. See section \ref{section_AppendixConcreteDynamics} for some more notes about symmetries in this paradigm.

		\subsubsection{Kleisli category and Cauchy surfaces}
		\label{section_ConcreteDynamicsKCategory}
		We now shift our attention to \textbf{free dynamical systems}, i.e. those in the form:
		\begin{equation}
			\mult{A}: \monad (\monad A) \epim \monad A
		\end{equation}
		The spaces of free histories $\monad A$ are canonically free dynamical systems, and their dynamical symmetries $h: \Automs{\CategoryC^\monad}{\mult{A}}$ are exactly the symmetries $h : \Automs{\CategoryC}{\monad A}$ that respect the free dynamical structure of the underlying physical system $A$. Section \ref{section_TimeFlows} of the appendix characterises propagators as dynamics of these systems, and thus we will also refer to them as \textbf{spacetimes}.\\

		Free dynamical systems form a full subcategory $\CategoryC_\monad$ of the Eilenberg-Moore category, isomorphic to the \textbf{Kleisli category}. There is a bijection between the morphisms $g: \mult{A} \rightarrow \mult{B}$ of free dynamical systems and morphisms of physical systems in the form $f: A \rightarrow \monad B$:  
		\begin{equation}
		\begin{array}{ccc}
			\text{Eilenberg-Moore}	& 					& \text{Kleisli} \\
			g:\mult{A} \rightarrow \mult{B} 			& \longrightarrow	& g \cdot \unit{A}\\
			f^\star \eqdef \mult{B} \cdot \monad f & \longleftarrow 	& f
		\end{array}
		\end{equation}
		where $f^\star$ is called the \textbf{Kleisli extension} of $f$. Thus morphisms of free dynamical systems can equivalently be seen as morphisms of physical systems (their \textbf{Kleisli form}), with the following composition law inherited from the Eilenberg-Moore category:
		\begin{equation}
			f_2 \circ^{Kleisli} f_1 \eqdef f_2^\star \cdot f_1 
		\end{equation}

		In particular, the dynamical symmetries $g : \Automs{\CategoryC^\monad}{\mult{A}}$ of a space of free histories $\monad A$ correspond bijectively to the morphisms $f : A \rightarrow \monad A$ that are iso in the Kleisli category:
		\begin{equation}
			 \left( (f^\star)^{-1}\cdot \unit{A} \right) \circ^{Kleisli} f = \id{A}^{Kleisli} \eqdef \unit{A}
		\end{equation}
		i.e. the morphisms $f: A \rightarrow \monad A$ s.t. $\mult{A} \cdot \monad f$ is invertible (which we'll denote by $f : \Automs{\CategoryC_\monad}{A}$). But what do these morphisms look like? First of all, note that if $f$ is the Kleisli form of a symmetry of a free dynamical system $\mult{A}$, then in particular it is (split) mono:
		\begin{equation}
			\label{equation_InvertibleKleisliForm}
			 (f^\star)^{-1} \cdot f = \unit{A}
		\end{equation}
		and hence it embeds $A$ as a faithful subsystem $A \stackrel{f}{\monom} \monad A$ of its space of free histories, with the automorphism $f^\star = \mult{A} \cdot \monad f : \Automs{\CategoryC}{\monad A}$ mapping the initial surface $\unit{A}: A \monom \monad A$ to it.\\

		To get an understanding of the action of $f^\star$ over the entire space we look once again at the mathematical theory of dynamical systems:
		\begin{equation}
		\begin{array}{ccccc|ccccc}
			f: & A & \rightarrow & \monad A = A \times \timeobj && \mult{A} \cdot \monad f: & \monad A & \rightarrow & \monad A \\
			& a & \mapsto & (g(a),t(a))&&						 & (a,\Delta t)    & \mapsto     & (g(a),t(a)+\Delta t)
		\end{array}
		\end{equation} 
		In this case, the map factors as an automorphism $g : \Automs{\CategoryC}{A}$ and a field of time translations $t: A \rightarrow \timeobj$: it is in fact what we'd usually call a \inlineQuote{Cauchy surface}.\\

		Informally, a Cauchy surface is a surface that intersects all maximal causal curves exactly once. Operationally, given a Cauchy surface $\Sigma$ any other Cauchy surface can then be obtained by dynamical symmetries, simply by deforming $\Sigma$ along the individual causal curves.\\

		Now two considerations from our conceptual framework come into play:
		\begin{enumerate}
			\item \textbf{Closure.} Since our system is closed in its evolution, all Cauchy surfaces will \inlineQuote{look like $A$}, i.e. they will be faithful subsystems $f : A \monom \monad A$.

			\item \textbf{Initial surface.} One of the operational requirements in our framework was the possibility of canonically setting initial value problems. As a consequence we already have a canonical Cauchy surface, the initial surface $\unit{A}: A \monom \monad A$. 
		\end{enumerate} 
		Thus we define a \textbf{Cauchy surface} as a faithful subsystem $f :  A \monom \monad A$ that can be  pulled back to the initial surface via some dynamical symmetry of the space of free histories:
		\begin{equation}
			\label{equation_CauchySurfaceDef}
			h^{-1} \cdot f = \unit{A} \text{ for some } h : \Automs{\CategoryC^\monad}{\mult{A}}
		\end{equation}

		As seen in eq'n \ref{equation_InvertibleKleisliForm}, every dynamical symmetry of the space of free history gives a Cauchy surface as its Kleisli form. But the opposite is also true: if $f$ is a Cauchy surface according to eq'n \ref{equation_CauchySurfaceDef}, then by uniqueness of inverse for $h$ in $\CategoryC^\monad$ we have that $h = f^\star$ and we recover eq'n \ref{equation_InvertibleKleisliForm}. We conclude that Cauchy surfaces are captured exactly by the automorphisms of spacetimes in the Kleisli category.
%% END SECTION - An operational notion of time %%

%% BEGIN SECTION - Time in symmetric monoidal categories %%
\newpage
\section{Time in symmetric monoidal categories}
	\label{section_TimeSymmetricMonoidal}

	\subsection{Time from change}
		\label{section_StrongMonads}

		\subsubsection{Strong monads}

		If physical systems have some distinguished compositional structure, encoded by a (symmetric) monoidal structure $(\CategoryC,\tensor,\tensorUnit)$, then it is interesting to ask whether there is a (physical) relation between the dynamics of individual systems and the dynamics of composite systems.\\  

		In \cite{CTh-Kock1970}\cite{CTh-Kock1972}\cite{LcC-Moggi1991}\cite{CTh-Moerdijk2002}, a \textbf{monoidal monad} is defined to be a monad $\monad$ on a symmetric monoidal category $(\CategoryC, \tensor, \tensorUnit)$ which comes with a natural family $m_{A,B}$ of morphisms relating the spaces of free histories of the individual systems to the space of free histories of their joint system:
		\begin{equation}
			m_{A,B}: \monad A \tensor \monad B \rightarrow \monad (A \tensor B)
		\end{equation}

		The definition of monoidal monad directly formalises our initial question. On the other hand, it can be shown that a monoidal monad is the same as a \textbf{commutative strong monad}, which comes with two natural transformations:
		\begin{equation}
		\label{equation_strenghtCostrenght}
		\begin{array}{cccc}
			t_{A,B}: &A \tensor \monad B &\rightarrow &\monad (A \tensor B)\\
			t'_{A,B}: &\monad A \tensor B &\rightarrow &\monad (A \tensor B)
		\end{array}
		\end{equation}
		Indeed one obtains $t_{A,B}$ as $m_{A,B} \cdot (\unit{A} \tensor \id{\monad B})$ and $t'_{A,B}$ by symmetry. There is a sense in which the transformations are associative, commute with each other, and respect $\tensorUnit$, $\unit{}$ and $\mult{}$: see the references for the 5 commutative diagrams that formalise those notions.\\

		\subsubsection{Foliation maps}

		We are interested in the equivalent formulation as (commutative) strong monad because these come with an evident natural \textbf{foliation} of the space of free histories:
		\begin{equation}
			\foliate{A} \eqdef \monad[\text{rightUnitor}_{A}] \cdot t_{A,\tensorUnit}: A \tensor \monad \tensorUnit \longrightarrow \monad (A \tensor \tensorUnit ) \isom \monad A 
		\end{equation}

		From these foliations we can define the \textbf{time object}, \textbf{time system} or simply \textbf{time}, to be the space of free histories $\timeobj \eqdef \monad \tensorUnit$ of the trivial system: $\foliate{A}$ can indeed be seen as a foliation of the space of free histories in same-time leaves, and naturality of foliation relates the \textbf{same-time action} $f \tensor \id{\timeobj}$ of a morphism $f$ to its lifting $\monad f$:
		\input{modules/diagrams/SymmetricMonoidalFramework/FoliationNaturality}
		We stress that the foliation does not, in general, need to cover the space of free histories, and that the leaves do not need to be faithful embeddings of $A$ into $\monad A$.

		\subsubsection{Concrete histories as curves parametrised by time}

		In section \ref{section_freeConcreteHistoriesGeneral} we introduced the notion of free history for a subsystem. For symmetric monoidal categories, there is a particular class of subsystems we are interested in, the \textbf{states} of a system:
		\begin{equation}
			\States{}{A} \eqdef \Hom{\CategoryC}{\tensorUnit}{A}
		\end{equation}
		Morphisms $f: A \rightarrow B$ have the usual pushforward action on states, given by:
		\begin{equation}
		\begin{array}{cccc}
			\pushforward{f}: &\States{}{A} &\rightarrow &\States{}{B} \\
						     & \psi 	   &\mapsto 	& f \cdot \psi
		\end{array}
		\end{equation}

		The free histories for the states of a system are the reason we refer to $\timeobj$ as the time of our theory, as a concrete history $\concretehistory{\alpha}{\psi}$ of a state $\psi : \States{}{A}$ is a curve in $A$, parametrized by time, which at the \textbf{initial time} $\unit{\tensorUnit} : \States{}{\timeobj}$ goes through $\psi$: 
		\begin{equation}
		\hbox{\input{./modules/pictures/FreeHistories3bis}}
		\end{equation}
		Furthermore, the canonical evolution gives an action of time on itself, which makes time into a commutative monoid $(\timeobj, \timeaction, \unit{\tensorUnit})$ encoding \textbf{time translation}:
		\begin{equation}
			\timeaction := \mult{\tensorUnit} \cdot \foliate{\timeobj}: \timeobj \tensor \timeobj  \longrightarrow  \timeobj
		\end{equation}
		It can be shown\footnote{It is a direct consequence of section \ref{section_dynamicsActionsTimePhysicalSystems}.} that concrete histories are compatible with this time translation. The full construction can be found in section \ref{section_ActionTimeItself} of the appendix.\\

		In conclusion, dynamics that are compatible with the compositional structure of physical systems always come with a canonical notion of time $(\timeobj, \timeaction, \unit{\tensorUnit})$, making concrete histories into time-parametrised curves. On the other hand, there need not always be a natural way to faithfully see dynamics as actions of time on physical systems, but this is a topic for the next section. 

	\subsection{Change from time}

		\subsubsection{Uniform monads}
		\label{section_UniformMonads}

		Any monoid  induces a canonical monad, which we will call a \textbf{uniform monad}, as the endofunctor
		\begin{equation}
		\begin{array}{cccc}
			\emptyArg \tensor \timeobj: & \CategoryC & \rightarrow & \CategoryC\\
					& A & \mapsto     & A \tensor \timeobj\\
					& f & \mapsto 	  & f \tensor \id{\timeobj}

		\end{array}
		\end{equation} 
		along with unit $\bar{\unit{}}_A := \id{A} \tensor \unit{\tensorUnit}: A \monom A \tensor \timeobj$ and multiplication $\bar{\mult{}}_A := \id{A} \tensor \timeaction: A \tensor \timeobj \tensor \timeobj \epim A \tensor \timeobj$; the triangle identities and square identity follow from the results on $\timeaction$ proven in section \ref{section_ActionTimeItself} of the appendix. Equivalently, uniform monads can be characterised as commutative strong monads where the foliations are all isomorphisms.\\

		The dynamics for a uniform monads are exactly the actions $\alpha: A \tensor \timeobj \epim A$ of time, as an internal monoid, on physical systems. Thus, at least for uniform monads, dynamics are determined by time.

		\subsubsection{Dynamics as actions of time on physical systems}
		\label{section_dynamicsActionsTimePhysicalSystems}		 

		If the monoid $(\timeobj, \timeaction,\unit{\tensorUnit})$ is the time  of some commutative strong monad $\monad$, then we shall refer to the monad $(\emptyArg \tensor \timeobj, \bar{\unit{}}, \bar{\mult{}})$ as the \textbf{associated uniform monad} for $\monad$. It is not hard to prove that foliation is not just a natural transformation, but in fact a morphism of monads:
		\begin{equation}
			\foliate{}: (\emptyArg \tensor \timeobj, \bar{\unit{}}, \bar{\mult{}}) \rightarrow (\monad, \unit{}, \mult{})
		\end{equation} 

		In particular, algebras $\alpha$ for the monad $\monad$ will be pulled back, via the foliation maps, to algebras $\bar{\alpha}$ for the associated uniform monad which will yield the same concrete histories for all states:
		\begin{equation}
		\hbox{\input{./modules/pictures/LiftingAlgebras3nobox}}
		\end{equation}
		Thus, foliation maps always allow us to see dynamics as actions of time on physical systems, by lifting them to dynamics for the associated uniform monad. See section \ref{section_AppendixUniformMonads} of the appendix for the various proofs.

		\subsubsection{Epically strong monads}
		\label{section_StrongerMonads}

		Foliation maps always make it possible to see dynamics as actions of time, and concrete histories of states are always invariant, but in general the lifting of dynamics need not be a faithful process, and histories of arbitrary subsystems may be altered (unless the theory has enough states).\\

		We will call a monad \textbf{epically strong} if the foliations are all \textbf{covering}, i.e. epimorphisms: 
		\begin{equation}
			\forall A . \; \foliate{A}: A \tensor \timeobj \epim \monad A
		\end{equation}
		Intuitively, this means that the entire space of free histories can be foliated, not necessarily without \inlineQuote{singularities}, by time-indexed leaves isomorphic to the physical system $A$. Also, this is equivalent to requiring that $t_{A,B}$ are all epimorphisms, because of the following identity:\footnote{Together with the fact that $[f \cdot g \text{ epi} \imply f \text{ epi}]$ in general.}
		\begin{equation}
			t_{A,B} \cdot (\id{A} \tensor \foliate{B}) = \foliate{A \tensor B} 
		\end{equation} 

		Dynamics can always be faithfully lifted actions of time on physical systems:
		\begin{equation}
		\begin{array}{cccc}
			\emptyArg \cdot \foliate{A}: &\Hom{\CategoryC}{\monad A}{\emptyArg} &\monom &\Hom{\CategoryC}{A \tensor \timeobj}{\emptyArg}\\
			&f:\monad A \rightarrow X & \mapsto & f \cdot \foliate{A}: A \tensor \timeobj \rightarrow X
		\end{array}
		\end{equation}
		Not all such actions, on the other hand, need be dynamical: the foliation map can be seen as imposing constraints\footnote{Especially if it is regular epi, see \cite{CTh-Borceux1}}, and the actions of time on a system that correspond to dynamics are exactly those that respect those constraints, i.e. factor through the foliation.\\

		In conclusion, dynamics that respect the compositional structure of physical systems\footnote{Aka given by monoidal/commutative strong monads.} come with a canonical notion of time, the space of free histories of the trivial system, and can always be seen as actions of time on physical systems by lifting them to dynamics for the associated uniform monad.\\

		Concrete histories of states are always invariant, but in order to guarantee the lifting to be a faithful process one has to require the original monad to be epically strong. It is no coincidence that most dynamics in physical theories seem to be described by epically strong monads: the latter are exactly those for which it is equivalent to see dynamics as actions of time on systems, possibly imposing some constraint on the actions by hand.

%% END SECTION - Time in symmetric monoidal categories %%

%% BEGIN SECTION - The space of concrete histories %%
\section{The space of concrete histories}
	\label{section_SpaceConcreteHistories}

	\subsection{State spaces}
		In physics, the most common approach to dealing with histories of a system is to consider histories of states under specific dynamics, rather than free histories: this section will be dedicated to reconciling our approach with the path-space approaches. To start off, recall that the \textbf{states} of a system $A$ are given by the hom-set:
		\begin{equation}
			\States{}{A} \eqdef \Hom{\CategoryC}{\tensorUnit}{A}
		\end{equation}
		and the states of all systems arrange themselves into the coslice category $\CosliceCat{\CategoryC}{\tensorUnit}$, the category of elements of the covariant hom-functor $\Hom{\CategoryC}{\tensorUnit}{\emptyArg}$; if the latter is faithful, i.e. if states $\tensorUnit \rightarrow A$ separate all parallel pairs of maps $f,g: A \rightarrow B$ (for all $A,B$ and all $f,g$), then we will say that the theory / category has \textbf{enough states}. From now on we will assume to have enough states; also, we will assume static systems $\counit{A}: A \tensor \timeobj \epim \timeobj$ exist, making the Eilenberg-Moore forgetful functor into a bundle $U: \CategoryC^\monad \epim \CategoryC$ of categories (see section \ref{section_operationalStructureTime} of the appendix for more details on static systems).\\

		The \textbf{state spaces} $\States{}{A}$ of physical systems $A : \CategoryC$ arrange themselves into a concrete category $\Hom{\CategoryC}{\tensorUnit}{\CategoryC}$, the \textbf{category of state spaces}, with morphisms given by 
		\begin{equation}
			F: \States{}{A} \rightarrow \States{}{B} \iff F = f \cdot \emptyArg \text{ with } f:A \rightarrow B
		\end{equation}
		Under the assumption of enough states, is immediate to check that $\CategoryC \isom \Hom{\CategoryC}{\tensorUnit}{\CategoryC}$: this shows that one can equivalently work with the original category of physical systems or with the corresponding category of state spaces. We will also assume to be working with uniform monads.\\

	\subsection{Spaces of concrete histories}
		Concrete histories are morphisms from time to physical systems, i.e. they live in $\Hom{\CategoryC}{\timeobj}{\CategoryC}$, but in fact more is true. It is immediate to see that concrete histories are $\timeobj$-module homomorphism $\concretehistory{\alpha}{\Psi_0}: \timemult \rightarrow \alpha$, and on the other hand one can show that all such homomorphisms\footnote{Satisfying $\Psi \cdot \timemult = \alpha \cdot (\Psi \tensor \id{\timeobj})$.} $\Psi: \timemult \rightarrow \alpha$ arise as concrete histories:
		\begin{equation}
			\Psi = \Psi \cdot \timemult \cdot (\timeunit \tensor \id{\timeobj}) = \alpha \cdot \left((\Psi \cdot \timeunit) \tensor \id{\timeobj}\right) = \concretehistory{\alpha}{\Psi \cdot \timeunit} 
		\end{equation}
		Thus concrete histories are not just the transformations of time, as the physical system $\timeobj$, into a physical system $A$, but in fact they coincide with the transformations of time, as the dynamical system $\timemult: \timeobj \tensor \timeobj \epim \timeobj$, into a dynamical system $\alpha: A \tensor \timeobj \epim A$, allowing the following neat definition for the \textbf{space of concrete histories} of a dynamical system:
		\begin{equation}
			\ConcreteHistories{}{\alpha} \eqdef \Hom{\CategoryC^\monad}{\timemult}{\alpha}
		\end{equation}
		It is possible to show \cite{StefanoGogioso-MonadicTimeQuantumTheory} that this formulation of concrete histories as module homomorphism is equivalent, in the case of Quantum Mechanics, to Shr\"odinger's equation.\\

		If $\alpha, \beta : \CategoryC^\monad$ are two dynamics and $f : \alpha \rightarrow \beta$ is a dynamical transformation\footnote{Satisfying $f \cdot \alpha = \beta \cdot (f \tensor \id{\timeobj})$.}, then $f$ can be lifted to a morphism between spaces of concrete histories for specific dynamics (forming a category $\ConcreteHistoriesEMCategory{\CategoryC}$, just like algebras form the Eilenberg-Moore category $\CategoryC^\monad$):
		\input{modules/diagrams/SymmetricMonoidalFramework/LiftingToMorphismOfConcreteHistories}
		and vice versa it's easy to prove, since our theory has enough states, that any such morphism comes from a dynamical transformation in this way. Thus we can define an isomorphism of categorical bundles\footnote{Technically, to guarantee the Eilenberg-Moore category is bundle over the original category one may have to require the existence of static systems, or equivalently that time can be discarded. This is not really relevant though, see section \ref{section_operationalStructureTime} of the appendix for more details about this.}, linking dynamics and spaces of concrete histories for them (see diagram \ref{diagrams_SymmetricMonoidalFramework_IsomEMBundles} p.\pageref{diagrams_SymmetricMonoidalFramework_IsomEMBundles}).\\

		We showed earlier that we can equivalently work with physical systems or with their state spaces: diagram \ref{diagrams_SymmetricMonoidalFramework_IsomEMBundles} extends this claim by showing that working with physical systems, dynamics and spaces of free histories is the same as working with state spaces and spaces of concrete histories. We conclude that the more common, static, \inlineQuote{concrete} approach to spaces of histories is equivalent to the monadic, dynamic, \inlineQuote{free} approach presented in this work. Section \ref{section_AppendixSpacesConcreteHistories} of the appendix contains a couple of additional remarks.\\

		\input{modules/diagrams/SymmetricMonoidalFramework/IsomEMBundles}
		
%% END SECTION - The space of concrete histories %%

%% BEGIN SECTION - Conclusions and future work %%
\section{Conclusions and future work}
	\label{section_Conclusions}

	In section \ref{section_AppendixSpacesConcreteHistories}, we have presented a monadic framework formalising an operational notion of dynamics in physical theories seen as the setting and evolution of initial value problems. We have identified in the Eilenberg-Moore category the natural environment for dynamical systems, and have defined Cauchy surfaces abstractly as automorphisms in the Kleisli category.\\

	In section \ref{section_TimeSymmetricMonoidal}, we have proven the main claim of the work, vindicating the Aristotelian view that time and change are defined by one another. We have shown that dynamics which respect the compositional structure of physical systems always define a canonical notion of time, and on the other hand that they can always be seen (not necessarily faithfully) as actions of time on physical systems. We have furthermore given an exact characterisation of the conditions allowing this lifting of dynamics to actions of time to be carried out in a faithful manner.\\

	Finally, in section \ref{section_SpaceConcreteHistories}, we have constructed state spaces and the analogous to path spaces. We have shown that, under some physically reasonable conditions, the monadic dynamics approach is equivalent in applicability to the more common path space approaches to dynamics.\\

	The appendix briefly drafts applications to a number of theories. Upcoming work from the author will focus on the applications of monadic dynamics to categorical quantum mechanics \cite{StefanoGogioso-MonadicTimeQuantumTheory}. Future work will further investigate the application to classical and relativistic mechanics, the devised applications of propagators to network theory and flow in graphical calculi, the formulation of ergodic and chaos theory in the monadic framework.\\

	The author wishes to thank Bob Coecke, Aleks Kissinger, Amar Hadzihasanovic, Nicol\`{o} Chiappori and Sukrita Chatterji for useful feedback, discussions and support. Funding from EPSRC and Trinity College is gratefully acknowledged.
%% END SECTION - Conclusions and future work %%

\newpage
%% BEGIN SECTION - Applications %%
\section{Appendix A: Applications}
	\label{section_Applications}

	\subsection{Draft applications}
		
		\subsubsection{Mathematical theory of dynamical systems}
		\label{section_DynamicalSystemsApplications}

		The mathematical theory of dynamical systems (e.g. refer to \cite{Msc-ChaoticEvolutionStrangeAttractors}\cite{Msc-HandbookDynamicalSystems}) is embodied by the category $\DiffCategory$ of smooth manifolds and differentiable maps, with cartesian product as composition of systems. The monadic construction has been carried in section \ref{section_GeneralFramework}.\\

		\subsubsection{Pure state quantum mechanics}
		\label{section_QMApplications}

		The Categorical Quantum Mechanics programme (e.g. see \cite{CQM-seminal}\cite{CQM-QuantumPicturalism}), and related works (e.g. see \cite{CQM-DeepBeauty}\cite{CQM-NewStructures}\cite{CQM-HigherQT}), show how Quantum Theory can be understood abstractly in terms of symmetric monoidal categories. In particular we're interested in quantum computation and concrete simulation of finite-dimensional quantum systems, so we'll be working in the category $\fdHilbCategory$ of finite-dimensional complex Hilbert spaces and linear maps.

		\begin{enumerate}

			\item Complex Hilbert spaces come with a symmetric monoidal structure, with the notion of joint system given by the tensor product $\tensor$ and the trivial system given by the 1-dim space $\tensorUnit := \complexs$. States of a system $\SpaceH$ are the morphisms $I \rightarrow \SpaceH$.

			\item The space of free histories for a system is given by $\monad \SpaceH := \SpaceH \tensor \timeobj$, with $\timeobj$ any fixed N-dim space (a clock with N positions).

			\item There is a distinguished basis $(\ket{t_n})_{n : \integersMod{N}}$ for $\timeobj$, with $N$ the length of time. The group $\integersMod{N}$ induces a monoid structure $(\timeobj, \mult{},\unit{})$ on $\timeobj$ by:
			\begin{equation}
			\begin{array}{clcl}
				\unit{}: 	& I & \rightarrow	& \timeobj 	\\
							& 1	& \mapsto		& \ket{t_0}	
			\end{array}
			\end{equation}
			\begin{equation}
			\begin{array}{clcl}
				\mult{}: 	& \timeobj \tensor \timeobj & \rightarrow	& \timeobj 	\\
							& \ket{t_n} \tensor \ket{t_m}	& \mapsto		& \ket{t_{n+m}}	
			\end{array}
			\end{equation}
			
			\item This yields the initial surface and canonical evolution. In fact this is a group structure, with time inversion given by 
			\begin{equation}
			\begin{array}{clcl}
				\timeinversion{}: 	& \timeobj & \rightarrow	& \timeobj 	\\
							& \ket{t_n}	& \mapsto		& \ket{t_{-n}}	
			\end{array}
			\end{equation}

			\item The lifting of maps is given by same-time action, similarly to the previous example:
			\begin{equation}
			\begin{array}{clcl}
				\monad f := f \tensor \id{\timeobj} : 	& \monad A 	&\rightarrow & \monad B\\
				& \ket{\varphi} \tensor \ket{t_n}	&\mapsto & (f\ket{\varphi}) \tensor \ket{t_n}
			\end{array}
			\end{equation}

		\end{enumerate}

		The detailed construction, and its applications to time in quantum mechanics, can be found in the upcoming \cite{StefanoGogioso-MonadicTimeQuantumTheory}: as an example of application, the construction can be used to understand time / energy duality in quantum mechanics in terms of the notion of \textbf{strong complementarity} introduced by \cite{CQM-QuantumClassicalStructuralism}. The issues with extending this approach to infinite time are also covered in \cite{StefanoGogioso-MonadicTimeQuantumTheory}, and are related to the work presented in \cite{CQM-HStarAlgebras}.

		\subsubsection{Quantum measurements}
		\label{section_QuantumMeasurementsApplications}

		The operational understanding of quantum measurements is one of the many faces of the Categorical Quantum Mechanics programme, and is presented in \cite{CQM-QuantumClassicalStructuralism}\cite{CQM-QuantumMeasuNoSums}\cite{CQM-QCSnotes}. The monadic dynamic framework can be used to characterise \textbf{non-demolition measurements}\footnote{In the sense of \cite{CQM-QCSnotes}.} as the concrete dynamics in mixed-state quantum mechanics for a particular notion of \textbf{1-step} time.

		\subsubsection{Classical and relativistic mechanics}
		\label{section_ClassicalRelativisticMechanicsApplications}

		The work of Jean-Marie Souriau \cite{ClM-SouriauGeometrieRelativite}\cite{ClM-SouriauStructureSystemesDynamiques} on the \textbf{evolution space} of a mechanical system, as presented by \cite{ClM-MarleSeminar}, fits perfectly into the monadic framework, and will be the subject of future work. 

		\subsubsection{Notions of time in computation}
		\label{section_ComputationApplications}

		As mentioned in section \ref{section_RelatedWork}, monads have a well-established role in the abstract modelling of computation, and references \cite{LcC-Moggi1991}\cite{LcC-Plotkin2002} give a number of constructions of interest. Each of those monads comes with a particular notion of time, and computations can be characterised as morphisms of dynamical systems. Particularly interesting is the notion of time for interactive input, which gives a model of branching time of somewhat Everettian flavour.

		\subsubsection{The dynamics of exploration}
		\label{section_DynamicsOfExploration}

		In order to better clarify the breadth of application of the framework, we briefly consider the following 3 views of spacetime in field theory, in order of increasing relativism:

		\begin{enumerate}
			\item the value of a field over the entire spacetime is known;
			\item the value of a field over the entire space at some initial time is known, and evolved through time: its space component is part of the physical system, its time component is encoded by the concrete dynamic chosen;
			\item the value of a field at a point is known, and the field is evolved through time and explored through space: both the time component and the space component are encoded by concrete dynamics.
		\end{enumerate}

		The last point of view shows what we could call \inlineQuote{dynamics of exploration}: one has a setting which, when seen in its entirety, would be static, but is understood through the dynamics. A monad modelling this setting could for example be $\monad A := A \times X \times \timeobj$, where $A$ is the system encoding the possible values of the field at a point of spacetime, $X$ is the group of motions associated with the underlying space (encoding the exploratory dynamics) and $\timeobj$ is the time object (encoding the evolutionary dynamics).\\

		Applications of these \inlineQuote{dynamics of exploration} might include: 
		\begin{enumerate}
			\item the modelling of empirical and experimental scenarios, where a physical system is explored by manipulating several knobs;
			\item the modelling of games, where a starting position is known and the evolution of the game is an exploration of the tree of moves.
		\end{enumerate} 
		This exploratory point of view, and its applications, will be the subject of future work.

	\subsection{The operational structure of time}
		\label{section_operationalStructureTime}

		\subsubsection{Operations for time}
		We have just seen that the dynamics of the theory are nothing but actions of time on physical systems. We already know that time is a monoid $(\timeobj,\timeunit,\timemult)$, and the rest of this section will cover some additional structure for it that might be of interest.\\
 
		Firstly, we are interested to know whether there is a canonical notion of \textbf{static system}, i.e. whether each system comes with a natural \inlineQuote{static} dynamic which just discards time:  
		\begin{equation}
			\counit{A}: A \tensor \timeobj \epim A
		\end{equation}
		Naturality of $\counit{A}$ is what determines this static nature: it yields a well defined functor
		\begin{equation}
		\begin{array}{cccc}
			F: & \CategoryC & \rightarrow & \CategoryC^\monad \\
			& A & \mapsto & \counit{A}\\
			& (f:A \rightarrow B) & \mapsto & (f:\counit{A}\rightarrow\counit{B})
		\end{array}
		\end{equation}
		making our bundles $\counit{A}: A \tensor \timeobj \epim A$ into \textbf{trivial bundles}. We can understand the dynamic $\counit{A}$ as \inlineQuote{doing nothing} to the physical system $A$.\\

		It is immediate to see that an \textbf{erasure} operation $\timeerase: \timeobj \epim I$ for time will do the job by setting $\counit{A} := \id{A} \tensor \timeerase$; vice versa we can always recover an erasure operation by taking $\timeerase := \counit{\tensorUnit}$, and the static systems will necessarily be given by $\counit{A} := \id{A} \tensor \counit{\tensorUnit}$ (if we assume our theory has enough states\footnote{I.e. any two parallel maps $f,g: A \rightarrow B$ that coincide on all states of the domain $A$ are in fact equal; equivalently, the functor $\Hom{\CategoryC}{\tensorUnit}{\emptyArg}$ is faithful.}).\\

		This idea that the dynamics $\counit{A}$ are static is further confirmed by their action on free histories, which discards time leaving the initial value untouched:
		\begin{equation}
			\concretehistory{\counit{A}}{\psi} = \counit{A} \cdot \freehistory{\psi} = \psi \tensor \timeerase
		\end{equation}

		Secondly, in section \ref{section_TimeFlows} we will need time to be a \textbf{copiable} resource. By this we mean that we have an erasure operation $\timeerase$ which can be extended to a comonoid $(\timeobj, \timeerase, \timediag)$ satisfying the bialgebra law and duplicating the initial time.\footnote{For more details on the connections with copiability see e.g. \cite{CQM-StrongComplementarity}\cite{CQM-seminal}\cite{CQM-QuantumClassicalStructuralism}.} There is no need for all states $\tensorUnit \rightarrow \timeobj$ of time to be copiable like the initial time (states that do, though, are closed under $\timemult$).\\

		Thirdly, it is interesting to have a notion of \textbf{time inversion}, i.e. an involution $\timeinverse = \timeinverse^{-1}: \timeobj \rightarrow \timeobj$ that satisfies the Hopf law with the comonoid above. \\

		If physical systems form a dagger category, then we say that the dagger structure is \textbf{compatible} with time inversion, i.e. can be interpreted as abstracting time-reversal of transformations, if daggering the action of any time state (under some fixed dynamic $\alpha$) yields the same result as letting the inverse time state act (see eq'n \ref{equation_FixedTimeAction} for notation):
		\begin{equation}
			\restrict{\alpha}{\timeinverse t} = (\restrict{\alpha}{t})^\dagger \text{ for all } t: \tensorUnit \rightarrow \timeobj
		\end{equation}

		Finally, we would like to mention the role of richer time structures found in compact closed symmetric monoidal categories. If the group structure of invertible time can be enriched to a full Frobenius algebra, then it can be shown that strong complementarity (see e.g. \cite{CQM-StrongComplementarity}) plays the role of time / frequency duality\footnote{In fact, modulo an $\hbar$ conversion factor, it plays the role of time / energy duality, and can be similarly extended to position / momentum duality if the monad encodes the dynamics of spatial translation.}, and can be used to characterise Fourier transforms in terms of a change of basis. More details about this can be found in the aforementioned \cite{StefanoGogioso-MonadicTimeQuantumTheory}.

	\subsection{Propagators}
		\label{section_TimeFlows}

		How do we keep track of evolution of a state in time? The space of free histories gives us the perfect abstract tool to do that, as long as we're interested in knowing its history at states of time that can be copied. From now on, we will implicitly assume (unless explicitly stated otherwise) that time is commutative and copiable, and assumption that time is invertible will be made explicitly when needed; most of the results can be extended to the smooth case, but this will be the subject of separate, future work.\\  

		Given a dynamical system $\alpha: A \tensor \timeobj \epim A$, we define its (time-translationally invariant) \textbf{propagator} $\timeflow{\alpha}: A \tensor \timeobj \tensor \timeobj \epim A \tensor \timeobj$ to be the following morphism:
		\begin{equation}
			\hbox{\input{./modules/pictures/TranslationallyInvariantPropagatorTikZ}}
			\label{figure_timeFlowDef}
		\end{equation}	
		The propagator of $\alpha$ takes a system in state $\psi$ at time $t$, a (copiable) amount $\Delta t$ of time to evolve the system for, and returns the evolved system state $\alpha \cdot (\psi \tensor \Delta t)$ at the new time $\timemult \cdot (t \tensor \Delta t)$. It is worth mentioning that time-translation can be seen as a propagator $\mult{A} = \timeflow{\counit{A}}$ for systems coming with static dynamics.\\

		It is easy to see that propagators of dynamics are themselves dynamics: the initial time and the multiplication are always copied, and the fact that both $\alpha$ and $\timemult$ are algebras of the monads takes care of the rest; so propagators of dynamics for a system $A$ give us our first examples of non-free dynamics for $A \tensor \timeobj$.\\

		We now define the \textbf{restriction} of a dynamic $\alpha: A \tensor \timeobj \epim \timeobj$ to a time state $t:\States{}{\timeobj}$:
		\begin{equation}
			\label{equation_FixedTimeAction}
			\restrict{\alpha}{t} \eqdef \alpha \cdot (\id{A}\tensor t)
		\end{equation}
		Fixing any amount of time $\Delta t$ to evolve a dynamical $\alpha: A \tensor \timeobj \epim A$ system for, we can turn a propagator $\timeflow{\alpha}$ into an endomorphism $\dynamicGenerator{\timeflow{\alpha}}{\Delta t}: A \tensor \timeobj \rightarrow A \tensor \timeobj$; since we assumed time to be commutative, it can be proven that this endomorphism is in fact an endomorphism $\timemult \rightarrow \timemult$ of time. By taking $\alpha = \timemult$, one can further show that time is commutative if and only if all $\alpha$ and all $\Delta t$ give an endomorphism of time (conditional to the existence of enough copiable time states).

		Propagators give us a way of tracking the evolution of a system under a fixed, time invariant dynamic; there are plenty of applications, on the other hand, where more flexibility is needed, flow of circuits and time-varying Hamiltonians amongst them. Luckily there are more dynamics for the spaces of free histories yet to come.\\

		We will refer to dynamics $\beta: A \tensor \timeobj \tensor \timeobj \epim A \tensor \timeobj$ in general as \textbf{propagators}, reserving the term \textbf{propagators of dynamics}, or \textbf{time-translationally invariant propagators}, to dynamics for $A \tensor \timeobj$ in the form $\timeflow{\alpha}$. We can see a propagator $\beta$ equivalently as an endomorphism:
		\begin{equation}
			\dynamicGenerator{\beta}{\timegenerator} : A \tensor \timeobj \rightarrow A \tensor \timeobj
		\end{equation}

		For the purposes of modelling things like time-dependent Hamiltonians\footnote{We will stick to their discrete counterparts in this work.}, we will consider \textbf{Markovian} propagators, i.e. propagators that take the form:
		\begin{equation}
			\dynamicGenerator{\beta}{\timegenerator} = \psi \tensor t \mapsto U_t \psi \tensor (t + \timegenerator)
		\end{equation}

		Discarding time, we immediately have a time-dependent family of generators:
		\begin{equation}
		\begin{array}{cccc}
			\counit{A} \cdot \restrict{\beta}{\timegenerator} : &A \tensor \timeobj &\rightarrow &\timeobj \\
			& \psi \tensor t &\mapsto & U_t \psi
		\end{array}
		\end{equation}

		On the other hand, any morphism\footnote{The notation $(U_t)_t$ for morphisms $f:A\tensor \timeobj \rightarrow \timeobj$ can be justified by observing that, since the span of the generator is a prebasis, then time has enough states to separate morphisms from it, and thus knowing $f \cdot (\id{A} \tensor t)$ for all $t:\States{}{\timeobj}$, or even just for $t : \Span{\timegenerator}$, uniquely identifies $f$.} $(U_t)_t : A \tensor \timeobj \rightarrow \timeobj$ can be turned into an endomorphism $\Lambda$, a candidate family of generators for a Markovian propagator; there is, unfortunately, no guarantee that a propagator $\beta: A \tensor \timeobj \tensor \timeobj \epim A \tensor \timeobj$ will exist s.t. $\dynamicGenerator{\beta}{\timegenerator} = \Lambda$. The conditions under which this definition of propagators from families is possible, e.g. compact closure and the existence of bases, are covered more in detail in the aforementioned \cite{StefanoGogioso-MonadicTimeQuantumTheory}, which also deals with a notion of causality for propagators.\\

		A first application of propagators is the evolution operator for pure state quantum mechanics \cite{QTC-ModernQuantumMechanics} in the case of a time-dependent Hamiltonian $H(t)$: time dependence makes the underlying Hilbert space $\SpaceH$ a non-closed system for the purposes of modelling the dynamics, but we can close it by keeping track of time as well, i.e. by working in $\SpaceH \tensor \timeobj$.\\

		We'll take evolution from time $\ket{n} := n \timegenerator$ to time $\ket{n+1}$ to be given by the unitary:
		\begin{equation}
			\dynamicGenerator{U}{\timegenerator}\ket{n} := \exp[\frac{i}{\hbar} H\ket{n}] 
		\end{equation}
		where we're working on a basis $(\ket{n})_n$ for $\timeobj$ generated by $\ket{1} := \timegenerator$. Then the usual time-evolution operator on $\SpaceH$ can be defined as the propagator:
		\begin{equation}
		\begin{array}{cccc}
			U: &\SpaceH \tensor \timeobj \tensor \timeobj &\epim &\SpaceH \tensor \timeobj \\
			& \ket{\psi} \tensor \ket{n} \tensor \ket{m} & \mapsto & \prod\limits_{j=0}^{m-1}\dynamicGenerator{U}{\timegenerator}(\ket{n+j})
		\end{array}
		\end{equation}
		where the product is expanded right to left as $j$ increases.\\

		Other applications are left to future work: 
		\begin{enumerate}
			\item \textbf{Network theory.} Non time-translationally invariant, Markovian propagators can be used in the modelling of sequential electrical and signal flow circuits like those in \cite{CQM-CategoriesInControl}\cite{CQM-FongTransferReport}\cite{CQM-NetworkTheorySeminars}, while non-Markovian (and non-causal) propagators extend the modelling to loopy circuits. 

			\item \textbf{Other graphical calculi.} More in general one can model flow in graphical calculi associated with symmetric monoidal categories: for example, the author will be interested in working out the monadic dynamics for the Feynman diagrams calculi of \cite{CQM-FeynmanDiagramsCalculus}\cite{CQM-ProofNetsFeynmanDiagrams}.

			\item \textbf{Time travel.} Non-causal propagators can also be used to model some of the quantum mechanical time-travel proposals presented in \cite{QGR-JMAllenCTCs}.

			\item \textbf{General relativity.} Propagators can be used to generalise dynamics in non-singular spacetimes: this will be covered as part of the application of monadic dynamics to relativistic mechanics.
		\end{enumerate}
%% END SECTION - Applications %%

\newpage
\section{Appendix B: Beyond the text}
	\label{section_appendixBeyondText}

	\subsection{The operational approach to dynamics}
		\label{section_AppendixOperationalApproach}

		\subsubsection{Free dynamics}
		\label{section_AppendixFreeDynamics}
		The canonical evolution $\mult{A}$ is meant to encode the \textbf{free dynamics} of system $A$: by this we mean that it encodes the abstract compositional aspect of the evolution, i.e. it tells us what it means to \inlineQuote{evolve the system a bit, then evolve it some more}.\\

		Operationally, the concrete evolution of $A$, or more in general of its subsystems, proceeds as follows:
		\begin{enumerate}
			\item[1.] take some subsystem $d : \Subsys{}{A}$
			\item[2.] lift it to its free history $\freehistory{d}$
			\item[3.] map it to its concrete history $\concretehistory{\alpha}{d} = \alpha \cdot \freehistory{d}$
		\end{enumerate}
		Indeed if we take subsystem  $\id{A} : \Subsys{}{A}$ we recover the original dynamic $\alpha = \concretehistory{\alpha}{\id{A}}$.\\

		The process of \inlineQuote{evolving the system a bit, then evolving it a some more} is thus characterised as:
		\begin{enumerate}
			\item[1.] take some subsystem $d : D \rightarrow A$
			\item[2.] lift it to its free history $\freehistory{d}: \monad D \rightarrow \monad A$
			\item[3.] map it to its concrete history $\concretehistory{\alpha}{d} = \alpha \cdot \freehistory{d} : \monad D \rightarrow A$
			\item[4.] lift the latter to its free history $\freehistory{\concretehistory{\alpha}{d}} : \monad \monad D \rightarrow \monad A$
			\item[5.] map it to its concrete history $\concretehistory{\alpha}{\concretehistory{\alpha}{d}} = \alpha \cdot \freehistory{\alpha \cdot \freehistory{d}} : \monad \monad D \rightarrow A$
		\end{enumerate}

		The process above can equivalently be described in the following, more direct, way:
		\begin{enumerate}
			\item[1.] take the free history $H := \freehistory{\freehistory{d}} : \monad \monad D \rightarrow \monad \monad A$\\
			(i.e. $H : \FreeHistories{}{\monad A})$
			\item[2.] push it forward to the free history $\pushforward{\alpha}[H] = \freehistory{\concretehistory{\alpha}{d}}: \monad \monad D \rightarrow \monad A$\\
			(i.e. $\pushforward{\alpha}[H] : \FreeHistories{}{A}$)
			\item[3.] map it to the concrete history $\alpha \cdot \pushforward{\alpha}[H] = \concretehistory{\alpha}{\concretehistory{\alpha}{d}} : \monad \monad D \rightarrow A$
		\end{enumerate}
		The latter corresponds to the transformation $\monad \monad A \stackrel{\monad \alpha}{\longrightarrow} \monad A \stackrel{\alpha}{\longrightarrow} A$ applied to the subsystem $H$:
			\input{modules/diagrams/GeneralFramework/FreeEvolutionExplained}

		When we say that $\mult{A}$ encodes the free dynamics of a physical system $A$, we mean that the process above should yield the same results as the one below, which involves only one application of the dynamic $\alpha$ and leaves the abstract compositional aspects to the canonical evolution $\mult{A}$:
		\begin{enumerate}
			\item[1.] take the free history $H : \FreeHistories{}{\monad A}$, of subsystem $\freehistory{d} : \Subsys{}{\monad A}$
			\item[2.] map it to the concrete history $\mult{A} \cdot H : \Subsys{}{\monad A}$, encapsulating the abstract compositional aspect of dynamics of $A$
			\item[3.] map it to the concrete history $\alpha \cdot \mult{A} \cdot H : \Subsys{}{A}$
		\end{enumerate}

		Since the two processes should yield the same result, we will ask for the corresponding morphisms to coincide. This gives an operational characterisation of $\mult{A}$ as free dynamics for $A$ in the form of the following diagram:

\input{modules/diagrams/GeneralFramework/AlgebraFreeDynamicalStructure}
		For the mathematical theory of dynamical systems, this condition reads:
		\begin{equation}
			\evolve{\Delta t}{\evolve{t}{a_0}} = \evolve{t+\Delta t}{a_0}
		\end{equation}

		\subsubsection{Neutral dynamics}
		\label{section_NeutralDynamics}
		Being itself a concrete dynamic of a physical system, the canonical evolution of free histories must respect the relevant intial surface:
			\input{modules/diagrams/GeneralFramework/MonadTriangularIdentityNeutral}
		$\unit{\monad A}$ then provides a form of \textbf{neutral dynamics} on $A$, as exemplified by the mathematical theory of dynamical systems:
		\begin{equation}
			\evolve{0}{\evolve{t}{a_0}} = \evolve{t}{a_0}
		\end{equation}

		The canonical evolution also has to respect the free dynamics for the system $\monad A$:
			\input{modules/diagrams/GeneralFramework/MonadSquareIdentity}

		\subsubsection{Initial surface revisited}
		\label{section_InitialSurfaceRevisited}
		Not only the initial surface should allow us to specify initial value problems, but we also expect dynamics for a system $A$ to somehow get lifted to $\unit{A}: A \rightarrow \monad A$, i.e. to $A$ seen as an initial surface. Under our structure-inducing viewpoint, we translate this into the requirement that concrete histories for subsystems $d : D \rightarrow A$ of a physical system $A$ canonically correspond to concrete histories for $\unit{A} \cdot d$, their embedding as subsystems of the initial surface: 
			\input{modules/diagrams/GeneralFramework/MonadTriangularIdentityInitSurfExplained}
		We further strengthen\footnote{In fact most cases of interest have enough free histories and enough dynamics to separate morphisms from and to $\monad A$, in which case the two requirements are equivalent.} this to the following, more operational condition: 
			\input{modules/diagrams/GeneralFramework/MonadTriangularIdentityInitSurf}
		For the mathematical theory of dynamical systems, the condition above reads:
		\begin{equation}
			\evolve{\Delta t}{\evolve{0}{a_0}} = \evolve{\Delta t}{a_0}
		\end{equation}

	\subsection{Monadic formulation}
		\label{section_AppendixConcreteDynamics}

		\subsubsection{Eilenberg-Moore category and dynamical systems}
		Particularly interesting is the lifting of $\Automs{\CategoryC}{A}$, the group of physical symmetries of a system $A$. The symmetries $\phi: \Automs{\CategoryC}{A}$ that get lifted to symmetries $\phi: \Automs{\CategoryC^\monad}{\alpha}$ of a dynamical system $\alpha: \monad A \epim A$ are exactly the time-independent \textbf{dynamical symmetries}, i.e. the part of $\Automs{\CategoryC}{A}$ that remains unbroken under that particular dynamic of system $A$:
			\input{modules/diagrams/GeneralFramework/EMCategoryDynSymmetries}

		For the mathematical theory of dynamical systems\footnote{Not the general ones defined above, the smooth manifolds with 1-parameter groups of endomorphisms.} this condition reads as usual:
		\begin{equation}
			\phi \cdot \evolve{\Delta t}{\emptyArg} = \evolve{\Delta t}{\emptyArg} \cdot \phi \text{ for all } t:\timeobj
		\end{equation}

		while for time-dependent dynamical symmetries $\Phi: \monad A = A \tensor \timeobj \rightarrow A$ one gets:
		\begin{equation}
			\Phi_{t+\Delta t}[\emptyArg] \cdot \evolve{\Delta t}{\emptyArg} = \evolve{\Delta t}{\emptyArg} \cdot \Phi_{t}[\emptyArg] \text{ for all } t,\Delta t : \timeobj
		\end{equation}
		The double appearance of $\Delta t$ on the LHS means that time must have a separate identity from the system $A$, and that it must be possible to duplicate its relevant states: section \ref{section_operationalStructureTime} of the appendix explains how this condition can be formalised.\\

		Since we mentioned \textbf{symmetries}, it is worth remembering that they form a category $\Symmetries{\CategoryC}$, with objects automorphisms $\phi: A \rightarrow A$ of $\CategoryC$ and morphisms: 
		\begin{equation}
			f: (\phi: \Automs{\CategoryC}{A}) \rightarrow (\phi' : \Automs{\CategoryC}{B}) \iff f: A \rightarrow B \text{ and } \phi' \cdot f = f \cdot \phi
		\end{equation}

	\subsection{Time from change}

		\subsubsection{Action of time on itself}
		\label{section_ActionTimeItself}

\begin{figure}[h!]
		%\begin{center}
			\begin{subfigure}{.55\textwidth}
			\begin{center}
				\input{modules/pictures/TimeActionAssociativityTikZ}
				%\caption{Associativity for the action of time on itself.}
			\end{center}
			\end{subfigure}
			\begin{subfigure}{.40\textwidth}
			\begin{center}
				\input{modules/pictures/TimeActionUnitTikZ}
			\end{center}
			\end{subfigure}
			\caption{Associativity and unit laws for the action $\nu$ of time on itself.}
			\label{figure_TimeActionUnitAssociativityLaws}
		%\end{center}
		\end{figure}
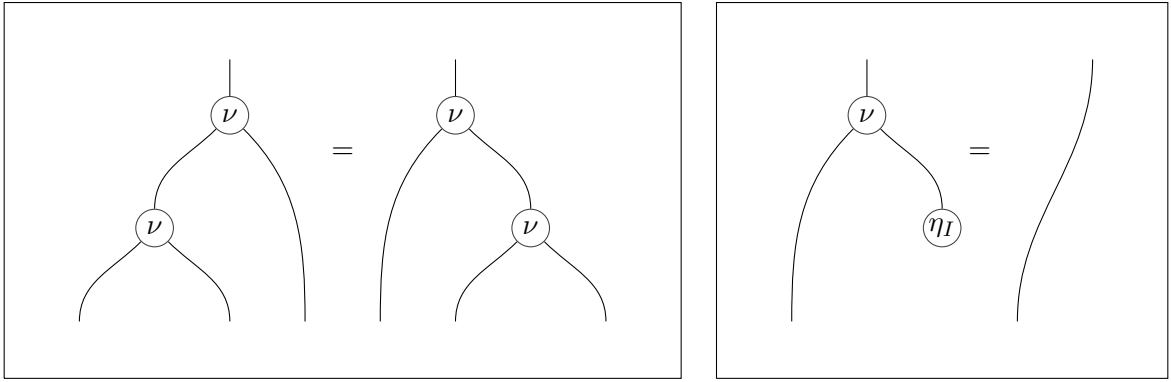

		The canonical evolution gives an action of time on itself as:
		\begin{equation}
			\timeaction := \mult{\tensorUnit} \cdot \foliate{\timeobj}: \timeobj \tensor \timeobj  \longrightarrow  \timeobj
		\end{equation}
		The action is associative and reduces to the identity on the initial time, as shown in figure \ref{figure_TimeActionUnitAssociativityLaws} (p.\pageref{figure_TimeActionUnitAssociativityLaws}).

		The proof of associativity is given by the following commuting diagram:
		\input{modules/diagrams/SymmetricMonoidalFramework/TimeActionAssociativity}
		where commutativity of the square labelled \inlineQuote{see below} is given by the following detail:
		\input{modules/diagrams/SymmetricMonoidalFramework/TimeActionAssociativityDetail}

		To prove the other unit law required to get a monoid $\timeaction: \timeobj \tensor \timeobj \epim \timeobj$, we need to use the $t'_{A,B}$ transformations from eq'n \ref{equation_strenghtCostrenght}. We start by defining the symmetric partner of $\foliate{A}$:
		\begin{equation}
			\cofoliate{A} \eqdef \monad[\text{leftUnitor}_{A}] \cdot t'_{A,\tensorUnit}: \timeobj \tensor A \longrightarrow \monad A 
		\end{equation}
		Then the definition of commutative strong monads guarantees\footnote{See \cite{CTh-Kock1972}, Prop. 2.1, Prop. 2.2 and Th. 2.3, remembering that $\foliate{\timeobj} = t_{\monad \tensorUnit,\tensorUnit}$.} that the following two definitions for the action of time on itself are consistent:
		\begin{equation}
		\begin{array}{cc}
			\timeaction := & \mult{\tensorUnit} \cdot \foliate{\timeobj} \\
			\timeaction := & \mult{\tensorUnit} \cdot \cofoliate{\timeobj}
		\end{array}
		\end{equation}
		The other unit law follows immediately by symmetry, and we have an associative monoid $\timeaction: \timeobj \tensor \timeobj \epim \timeobj$ with bilateral unit $\unit{\timeobj}$:
		\begin{equation}
			\nu \cdot (\id{\timeobj} \tensor \unit{\timeobj}) = \id{\timeobj} = \nu \cdot (\unit{\timeobj} \tensor \id{\timeobj})
		\end{equation}

	\subsection{Change from time}

		\subsubsection{Dynamics as actions of time on physical systems}
		\label{section_AppendixUniformMonads}

		Foliation is not just a natural transformation, but in fact a morphism of monads:
		\begin{equation}
			\foliate{}: (\emptyArg \tensor \timeobj, \bar{\unit{}}, \bar{\mult{}}) \rightarrow (\monad, \unit{}, \mult{})
		\end{equation}  
		where we defined $\bar{\unit{}}_A := \id{A} \tensor \unit{\tensorUnit}$ and $\bar{\mult{}}_A := \id{A} \tensor \timeaction$.\\

		Indeed by naturality of $t$ (and thus of foliation) it's immediate to see that:
		\input{modules/diagrams/SymmetricMonoidalFramework/FoliationRespectsUnit}

		Furthermore by associativity of $t$ one gets the following, where the two morphisms on the top coincide because of naturality of foliation:
		\input{modules/diagrams/SymmetricMonoidalFramework/FoliationRespectsMult}

		Going back to the pulled back algebras $\bar{\alpha} = \alpha \cdot \foliate{A}$ from \ref{diagrams_SymmetricMonoidalFramework_LiftingAlgebrasToUniformMonad}, we need to show that they allow to set initial value problems (from now on IVPs, see section \ref{section_InitialValueProblems}) for the associated uniform monad $(\emptyArg \tensor \timeobj, \unit{\tensorUnit}, \timeaction)$, and respect its free dynamics (see section \ref{section_FreeDynamics}).\\

		The possibility of setting IVPs follows from the fact that $t$ (and thus foliation) respects $\unit{}$, and that $\alpha$ allows for IVPs to be set for $\monad$: 
		\input{modules/diagrams/SymmetricMonoidalFramework/LiftedAlgebrasIVP}

		The free dynamics follow by naturality of $\foliate{}$, the free dynamics for $\alpha$, the fact that $t$ respects $\mult{}$, and the fact that $t$ is associative, natural and respects the unitors:
		\input{modules/diagrams/SymmetricMonoidalFramework/LiftedAlgebrasFreeDynamics}
		The proof of commutativity for the upper left quadrant goes by associativity and naturality, similarly to that of diagram \ref{diagrams_SymmetricMonoidalFramework_TimeActionAssociativity} (p.\pageref{diagrams_SymmetricMonoidalFramework_TimeActionAssociativity}).\\

		As a final note, uniform monads are particularly nice to work with, as their structure is completely determined by the monoid encoding the action of time on itself, and there is a clear-cut separation between the transformations / subsystems of the original system and its free dynamical structure. Furthermore, it is easy to see that uniform monads are commutative\footnote{And therefore closed under composition (whereas general monads are known not to be).}, and can be used to combine multiple independent notions of dynamics together.

	\subsection{The space of concrete histories}
		\subsubsection{Spaces of concrete histories}
		\label{section_AppendixSpacesConcreteHistories}
		As an aside, we note that free histories in $\FreeHistories{}{A}$ are exactly those concrete histories for the free dynamic $\id{A} \tensor \timemult: A \tensor \timeobj \tensor \timeobj \epim A \tensor \timeobj$ that take their initial value in $A$, i.e. that factor through the unit $\unit{A}: A \monom A \tensor \timeobj$ at the initial time $\timeunit$.\\

		Finally, just like the free histories perspective is monadic in nature, the concrete histories perspective has a comonadic flavour to it, and lacks the dynamical character of the former. The morphisms $\concretehistory{\alpha}{\emptyArg}$ mapping states to the associated concrete histories are sections of the following bundle:
		\begin{equation}
		\label{equation_tangentBundle}
			\emptyArg \cdot \timeunit: \ConcreteHistories{}{A} \epim \States{}{A}
		\end{equation}
		and are in fact the coalgebras of a certain comonad on the \inlineQuote{higher path spaces} $\Hom{\CategoryC}{\timeobj^{\tensor n}}{A}$.

\newpage
\section{Appendix C: Conventions}
	\label{section_appendixStyle}

	New definitions are introduced by \textbf{boldface}, while proofs are integrated with the narrative. Some of the notational conventions might also be slightly unusual:
	\begin{enumerate}
		
		\item \textbf{Definitions.} We use $\eqdef$ for permanent definitions, $:=$ for temporary definitions (usually limited to the section), and $\equiv$ for notational / symbolic equivalence.
		
		\item \textbf{Composition.} Instead of the mathematically more common $\circ$, we use the lighter $\cdot$ to denote composition of morphisms (except when we need to disambiguate, e.g. for Kleisli composition). 
		
		\item \textbf{Wild-cards.} We use the wild-card $\emptyArg$ to denote functional arguments left empty.
		
		\item \textbf{Typing and belonging.} We freely confuse typing and belonging, and we use the colon notation $:$ instead of the set-theoretic $\in$. The application will usually be clear from the context, and the following guidelines should cover all cases of interest:
		\begin{enumerate}
			\item[a.] if $A$ is a set, then $a : A$ means that $a$ is an element of $A$;
			\item[b.] if $A$ is more in general an object of a monoidal category $(\CategoryC, \tensorUnit, \tensor)$, then $a: A$ means that $a$ is a state of $A$, i.e. $a : \tensorUnit \rightarrow A$;
			\item[c.] if $A,B$ are objects of a category $\CategoryC$, then $f: A \rightarrow B$ means, as usual, that $f$ is a morphism from $A$ to $B$, or equivalently the $f:\Hom{\CategoryC}{A}{B}$ in the set-belonging notion;
			\item[d.] if $\CategoryC$ is a category, then $A : \CategoryC$ means that $A$ is an object of $\CategoryC$
		\end{enumerate}

		\item \textbf{Monomorphisms and epimorphisms.} Following standard categorical notation, we denote monomorphisms by $\monom$ and epimorphisms by $\epim$.

		\item \textbf{Subobjects and subsystems.} We use the nomenclature \inlineQuote{subsystems of $A$} to denote \underline{all} morphisms $D \rightarrow A$, not just the monomorphisms (usually referred to as subobjects): this choice follows the view that a subsystem of $A$ is in general a way of seeing another system $D$ in $A$, and that this process need be faithful in turning subsystems of $D$ into subsystems of $A$.

	\end{enumerate}
	Other clashes with standard notation will be clarified in due course. 

\newpage
\section{Appendix D: Discussion of related work}
	\label{section_RelatedWork}

	The general understanding that physical theories can be turned into categories of systems and transformations has a number of sources of inspiration, but is perhaps best presented by the collected works \cite{CQM-DeepBeauty}\cite{CQM-NewStructures}\cite{QFT-DoeringIsham4}. The operational characterisation of dynamics is more affine to the spirit of categorical quantum mechanics and quantum theory \cite{CQM-seminal}\cite{CQM-QuantumPicturalism}\cite{CQM-QCSnotes}, while the idea of internalising time by physical simulation was inspired by \cite{QGR-QuantumGraphenity}.\\

	Only an elementary understanding of category theory is required, as most of the constructions will be carried out explicitly: \cite{CTh-Borceux1}\cite{CTh-Borceux2} will do well as references. The theory of monads, particularly in the context of symmetric monoidal categories, is exposed in the seminal works \cite{CTh-Kock1970}\cite{CTh-Kock1972}\cite{CTh-Street2002}\cite{CTh-Moerdijk2002}\cite{CTh-Street1972}. Furthermore, the 2-categorical nature of monads suggests that the higher quantum theory programme of \cite{CQM-HigherQT} will play a significant role in the application of monadic dynamics to quantum theory.\\

	The mathematical theory of dynamical systems \cite{Msc-HandbookDynamicalSystems} will provide a wealth of examples for the initial construction, a germ of the free histories idea for this theory having already appeared\footnote{Although from a completely different perspective and with different motivations.} in \cite{Msc-SymbolicDynamics}; applications to ergodic and chaos theory \cite{Msc-ChaoticEvolutionStrangeAttractors} will be covered by future work. Certainly the main application at present, the dynamic of quantum theory is presented in the separate upcoming work \cite{StefanoGogioso-MonadicTimeQuantumTheory} by the author; a similar, but unrelated, (co)monadic construction appeared in \cite{CQM-QuantumMeasuNoSums}\cite{CQM-QuantumClassicalStructuralism} in the context of measurements. The time / energy duality refers to the complementary structures of \cite{CQM-StrongComplementarity}, while the issues associated with infinite time are related to the work in \cite{CQM-HStarAlgebras}.\\

	Our framework can be applied to classic and relativistic mechanics following the evolution space construction of \cite{ClM-SouriauGeometrieRelativite}\cite{ClM-SouriauStructureSystemesDynamiques}\cite{ClM-MarleSeminar}: this application will be the subject of future work. Furthermore, monads have already played a prominent role as abstract models of computation for the past 20 years, to the point that the authors of \cite{LcC-Plotkin2002} assert that \inlineQuote{\textit{Notions of computation determine monads}}:\footnote{\inlineQuote{\textit{But are not identified with monads}}, they continue.}: future work will explore the application of the monadic dynamics framework to some computational monads from \cite{LcC-Moggi1991}.\\

	The propagator fragment of the framework can be used to model flow in graphical calculi, in particular electrical and control flow networks \cite{CQM-NetworkTheorySeminars}\cite{CQM-CategoriesInControl}\cite{CQM-FongTransferReport}, quantum circuits \cite{CQM-QuantumPicturalism}, quantum mechanical time-travel \cite{QGR-JMAllenCTCs}, Feynman diagrams \cite{CQM-FeynmanDiagramsCalculus}\cite{CQM-ProofNetsFeynmanDiagrams} and time-dependent Hamiltonians\cite{QTC-ModernQuantumMechanics}: these applications will be the subject of future work. Finally, the construction of a space of concrete histories connects the monadic dynamics framework with the histories approach commonly used, for example, by path integral formulations \cite{QFT-FeynmanPathIntegral}\cite{QFT-FeynmanPathIntegralBook}.

\bibliographystyle{eptcs}
\bibliography{./bibliography/CategoryTheory,./bibliography/CategoricalQM,./bibliography/NonLocalityContextuality,./bibliography/QuantumComputing,./bibliography/ClassicalMechanics,./bibliography/LogicComputation,./bibliography/Gravitation,./bibliography/QFT,./bibliography/StatisticalPhysics,./bibliography/Misc,./bibliography/StefanoGogioso}

\end{document}

%% file: modules/packages.tex
\usepackage{fullpage}
\usepackage{graphicx}
\usepackage{amsfonts}
\usepackage{amsmath}  
\usepackage{amsthm}
\usepackage{amssymb}
\usepackage{mathtools}
\usepackage[applemac]{inputenc} 
\usepackage{fancyvrb}
\usepackage{multicol}
\usepackage{diagrams}
\usepackage{subcaption}

\usepackage{tikz}
\usetikzlibrary{arrows,shapes,snakes,backgrounds,positioning}

%% file: modules/macros.tex
	\numberwithin{theorem_c}{section}

	\theoremstyle{plain}

	\theoremstyle{definition}
	\newcommand{\emptyArg}{\,\underline{\hspace{6px}}\,}
	\newcommand{\inlineQuote}[1]{\textquotedblleft #1\textquotedblright}

	\newcommand{\naturals}{\mathbb{N}}
	\newcommand{\integers}{\mathbb{Z}}
	
	\newcommand{\reals}{\mathbb{R}}
	\newcommand{\complexs}{\mathbb{C}}
	
	\newcommand{\integersMod}[1]{\mathbb{Z}_{#1}}
	
	\newcommand{\inject}{\hookrightarrow}
	\newcommand{\restrict}[2]{\left. #1 \right\vert_{#2}}

	\newcommand{\eqdef}{\stackrel{def}{=}}
	
	\newcommand{\imply}{\Rightarrow}
	\newcommand{\suchthat}[2]{\{#1 \: \text{ s.t. } \: #2\}}

		\newcommand{\isom}{\cong}
		\newcommand{\epim}{\twoheadrightarrow}
		\newcommand{\monom}{\rightarrowtail}

		\newcommand{\tensor}{\otimes}
		\newcommand{\tensorUnit}{I}
		\newcommand{\cartesianTensor}{\times}
		\newcommand{\id}[1]{id_{#1}}

		\newcommand{\iso}{\cong}

		\newcommand{\Hom}[3]{\operatorname{Hom}_{\,#1}\left[#2,#3\right]}

		\newcommand{\Automs}[2]{\operatorname{Aut}_{\,#1}\left[#2\right]}
		
		\newcommand{\Subsys}[2]{\operatorname{Subsys}_{#1}[#2]}
		\newcommand{\States}[2]{\operatorname{States}_{#1}[#2]}
		\newcommand{\Symmetries}[1]{\Automs{#1}{\emptyArg}}

			\newcommand{\DiffCategory}{\operatorname{Diff}}

			\newcommand{\fdHilbCategory}{\operatorname{fdHilb}}

			\newcommand{\CategoryC}{\mathcal{C}}

		\newcommand{\CosliceCat}[2]{#2 \backslash #1} % #1 Category, #1 object

		\newcommand{\monad}{T}
		\newcommand{\unit}[1]{\eta_{#1}}
		\newcommand{\mult}[1]{\mu_{#1}}
			\newcommand{\evolve}[2]{\operatorname{evolve}_{#1}[#2]} 
			\newcommand{\timeobj}{\mathbb{T}}

			\newcommand{\freehistory}[1]{\operatorname{freehist}_{#1}}
			\newcommand{\FreeHistories}[2]{\operatorname{FreeHists}_{#1}[#2]}
			\newcommand{\pushforward}[1]{#1_\star}
			\newcommand{\pullback}[1]{#1^\star}
			\newcommand{\concretehistory}[2]{\operatorname{hist}^{#1}_{#2}} % #1 = concrete dynamic, #2 = subsystem
			\newcommand{\ConcreteHistories}[2]{\operatorname{Hists}_{#1}[#2]}
			\newcommand{\ConcreteHistoriesEMCategory}[1]{\prescript{H}{}{#1}}
			\newcommand{\timeinversion}[1]{i_{#1}}
			\newcommand{\foliate}[1]{\operatorname{foliate}_{#1}}
			\newcommand{\cofoliate}[1]{\operatorname{foliate'}_{#1}}
			\newcommand{\timeaction}{\nu}
			\newcommand{\timeunit}{\unit{}}
			\newcommand{\timemult}{\mult{}}
			\newcommand{\timediag}{\delta}
			\newcommand{\timeerase}{\epsilon}
			\newcommand{\timeinverse}{\timeinversion{}}
			\newcommand{\timegenerator}{1_\timeobj}

			\newcommand{\timeflow}[1]{[#1]}
			\newcommand{\dynamicGenerator}[2]{\restrict{#1}{#2}}

		\newcommand{\counit}[1]{\epsilon_{#1}}

	\newcommand{\ket}[1]{\vert #1 \rangle}

	\newcommand{\SpaceH}{\mathcal{H}}
	
	\newcommand{\Span}[1]{\operatorname{Span}[#1]}

	\newarrow{Dashembed}{>}{dash}{}{dash}{>}
	\newarrow{Straightline}{}{-}{}{-}{-}

	\tikzstyle{trivial}=
		[
		draw=none,
		fill=none,
		thick
		]
	\tikzstyle{filler}=
		[
		draw=none,
		fill=none,
		thick,
		inner sep=0pt,
		minimum size=0mm
		] 
	\tikzstyle{system}=
		[
		fill=none
		]
	\tikzstyle{morphism}=
		[
		inner sep=0mm,
		minimum height=5mm,
		minimum width=8mm,
		draw=black!75,
		trapezium, trapezium left angle=90
		]
	\tikzstyle{state}=
		[
		inner sep=0mm,
		minimum height=5mm,
		minimum width=8mm,
		draw=black!75,
		triangle, shape border rotate=180
		]
	\tikzstyle{spider}=
		[
		inner sep=0mm,
		minimum size=5mm,
		draw=black!75,
		circle
		]
	\tikzstyle{identity}=
		[
		inner sep=0pt,
		minimum size=0.1pt
		]
	\tikzstyle{boundary}=
		[
		draw=none,
		fill=none,
		inner sep=0pt,
		minimum size=5mm
		]

	\tikzstyle{rectangle}=
		[
		draw=black,
		fill=none %,
		] 
	\tikzstyle{timespace}=
		[
		draw=orange!75,
		fill=orange!25,
		thick
		] 	
	\tikzstyle{freqspace}=
		[
		draw=green!75,
		fill=green!25,
		thick
		] 	
	\tikzstyle{algebra}=
		[
		draw=red!75,
		fill=red!25,
		thick
		]

%% file: modules/diagrams/GeneralFramework/MonadUnitMultNaturality.tex
%%\hfill\\
\begin{diagram}[LaTeXeqno,righteqno,balanced,heads=littlevee] 
\label{diagrams_GeneralFramework_MonadUnitNaturality}
	\monad A			&	\rTo^{\monad f}		& 	\monad B				&&	\monad \monad A  	&	\rTo^{\monad\monad f}	& 	\monad \monad B			\\
	\uEmbed<{\unit{A}} &						&	\uEmbed>{\unit{B}}		&&	\dOnto<{\mult{A}} 	&						&	\dOnto>{\mult{B}}		\\
	A					&	\rTo_{f}			&	 B 						&&	\monad A			&	\rTo_{\monad f}		&	\monad B 		
\end{diagram}
%\hfill\\

%% file: modules/diagrams/GeneralFramework/AlgebraInitialSurface.tex
\begin{diagram}[LaTeXeqno,righteqno,balanced,heads=littlevee]
\label{diagrams_GeneralFramework_AlgebraInitialSurface}
\monad A			&						&		\\
\uEmbed<{\unit{A}}	&	\rdOnto^{\alpha}	&		\\
A					& 	\rTo_{\id{A}}		&	A 					
\end{diagram}

%% file: modules/diagrams/GeneralFramework/AlgebraFreeDynamicalStructure.tex
\begin{diagram}[LaTeXeqno,righteqno,balanced,heads=littlevee]
\label{diagrams_GeneralFramework_AlgebraFreeDynamicalStructure}
\monad \monad A 	&	\rOnto^{\monad \alpha} 	&	\monad A 				\\
\dOnto<{\mult{A}}	&							&	\dOnto>{\alpha}			\\
\monad A 			&	\rOnto_{\alpha} 		&	A 						
\end{diagram}

%% file: modules/diagrams/GeneralFramework/EMCategoryMorphisms.tex
\begin{diagram}[LaTeXeqno,righteqno,balanced,heads=littlevee]
\label{diagrams_GeneralFramework_AlgebraFreeDynamicalStructure}
\monad  A 		&	\rTo^{\monad f} &	\monad B 				\\
\dOnto<{\alpha}	&					&	\dOnto>{\beta}			\\
 A 				&	\rTo_{f} 		&	B 						
\end{diagram}

%% file: modules/diagrams/SymmetricMonoidalFramework/FoliationNaturality.tex
\begin{diagram}[LaTeXeqno,righteqno,balanced,heads=littlevee]
\label{diagrams_SymmetricMonoidalFramework_FoliationNaturality}
A \tensor \timeobj 		&	\rTo<{f \tensor \id{\timeobj}} 		&	B \tensor \timeobj 				\\
\dTo^{\foliate{A}}		&										&	\dTo_{\foliate{B}}		\\
\monad A 				&	\rTo>{\monad f}	 	 				&	\monad B 						
\end{diagram}

%% file: modules/pictures/FreeHistories3bis.tex
\begin{tikzpicture}[baseline=(current  bounding  box.center)]
    \path[use as bounding box] (-1,-0.5) rectangle (5.5,3);

		\node [style] (I) at (0,0) {$\tensorUnit$};
		\node [style] (TI) at (0,1.5) {$\timeobj$};
		\node [style] (H1) at (2.5,0) {$A$};
		\node [style] (TH) at (2.5,1.5) {$\monad A$};
		\node [style] (H2) at (4.5,0) {$A$};

		\draw [style=>->] (I) -- node [left] {$\unit{\tensorUnit}$} (TI);
		\draw [style=>->] (H1) -- node [left] {$\unit{A}$} (TH);

		\draw [style=->]  (I) -- node [below] {$\psi$} (H1);
		\draw [style=->]  (TI) -- node [above] {$\freehistory{\psi}$} (TH);

		\draw [style=->]  (TH) -- node [above] {$\alpha$} (H2);
		\draw [double equal sign distance]  (H1) -- node [below] {} (H2);
		\draw [style=->] [out=90, in=90]  (TI)  to node [above] {$\concretehistory{\alpha}{\psi}$} (H2);
\end{tikzpicture}

%% file: modules/pictures/LiftingAlgebras3nobox.tex
\begin{tikzpicture}[baseline=(current  bounding  box.center)]

		\node [style] (T1) at (0,1.5) {$\timeobj$};
		\node [style] (T2) at (0,3) {$\timeobj$};
		\node [style] (TH) at (3,1.5) {$\monad A$};
		\node [style] (HT) at (3,3) {$A \tensor \timeobj$};
		\node [style] (H) at (3,0) {$A$};

		\draw [double equal sign distance]  (T1) to node [left] {} (T2);
		\draw [style=->]  (T1) to node [above] {$\freehistory{\psi}$} (TH);
		\draw [style=->]  (T2) to node [above] {$\psi \tensor \id{\timeobj}$} (HT);
		\draw [style=->] [out=270,in=180] (T1) to node [above] {$\concretehistory{\alpha}{\psi}$} (H);
		\draw [style=->>]  (TH) to node [right] {$\alpha$} (H);
		\draw [style=->]  (HT) to node [right] {$\foliate{A}$} (TH);
		\draw [style=->>] [out=0, in=0, distance=3cm] (HT) to node [right] {$\bar{\alpha}$} (H);

\end{tikzpicture}

%% file: modules/diagrams/SymmetricMonoidalFramework/LiftingToMorphismOfConcreteHistories.tex
\begin{diagram}[LaTeXeqno,righteqno,balanced,heads=littlevee]
\label{diagrams_SymmetricMonoidalFramework_LiftingToMorphismOfConcreteHistories}
\ConcreteHistories{}{A}							& \rTo^{f \cdot \emptyArg}	& \ConcreteHistories{}{B} 						\\
\uEmbed<{\concretehistory{\alpha}{\emptyArg}}	& 							& \uEmbed>{\concretehistory{\beta}{\emptyArg}}	\\
\States{}{A}									& \rTo_{f \cdot \emptyArg}	& \States{}{B}							
\end{diagram}

%% file: modules/diagrams/SymmetricMonoidalFramework/IsomEMBundles.tex
\begin{diagram}[LaTeXeqno,righteqno,balanced,heads=littlevee]
\label{diagrams_SymmetricMonoidalFramework_IsomEMBundles}
					&& A \tensor \timeobj	& \rTo^{f \tensor \id{\timeobj}}	& B \tensor \timeobj 	&		& \ConcreteHistories{}{A}						& \rTo^{f \cdot \emptyArg}	 & \ConcreteHistories{}{B}		 && \\
\CategoryC^\monad 	&& \dOnto<{\alpha}		&									& \dOnto>{\beta}		&		& \uEmbed<{\concretehistory{\alpha}{\emptyArg}}& 									& \uEmbed>{\concretehistory{\beta}{\emptyArg}}		&& \ConcreteHistoriesEMCategory{\CategoryC}\\
					&& A 					& \rTo^{f}							& B 					&		& \States{}{A}									& \rTo^{f \cdot \emptyArg}	 & \States{}{B}					 && \\
\dOnto				&&  					& \dMapsto							& 			 			&\isom 	& 												& \dMapsto							& 							 			 			&& \dOnto\\
\CategoryC			&& A 					& \rTo>{f}							& B 					&		& \States{}{A}									& \rTo>{f \cdot \emptyArg}	 & \States{}{B}							 			 && \States{}{\CategoryC}\\
\end{diagram}

%% file: modules/pictures/TranslationallyInvariantPropagatorTikZ.tex
\begin{tikzpicture}[baseline=(current  bounding  box.center)]

		\node [boundary] (in_system) at (0, 0) {};
		\node [boundary] (in_time1) at (1, 0) {};
		\node [boundary] (in_time2) at (2, 0) {};

		\node [spider] (timediag) at (2, 1) {$\timediag$};

		\node [morphism, ultra thick] (dynamic) at (0, 2.5) {$\alpha$};
		\node [spider] (timemult) at (2, 2.5) {$\timemult$};

		\node [boundary] (out_system) at (0, 3.5) {};
		\node [boundary] (out_time) at (2, 3.5) {};

		\draw [system,ultra thick] [out=90, in=270] (in_system.130) to (dynamic.230);
		\draw [system] [in=225, out=90] (in_time1) to (timemult);
		\draw [system] (in_time2) to (timediag);

		\draw [system] [in=270, out=135] (timediag) to (dynamic.310);
		\draw [system] [in=315, out=45] (timediag) to (timemult);

		\draw [system,ultra thick] (dynamic) to (out_system);
		\draw [system] (timemult) to (out_time);

\end{tikzpicture}

%% file: modules/diagrams/GeneralFramework/FreeEvolutionExplained.tex
\begin{diagram}[LaTeXeqno,righteqno,balanced,heads=littlevee]
\label{diagrams_GeneralFramework_AlgebraFreeDynamicalStructure}
\monad \monad A 	&	\rTo^{\monad \alpha} 		&	\monad A 						&	\rTo^{\alpha}				&	A											\\
\uTo<{H}			&								&	\uTo<{\pushforward{\alpha}[H]}	&								& \uTo{\alpha \cdot \pushforward{\alpha}[H]}	\\
\monad \monad D 	&	\rTo_{\id{\monad\monad D}} 	&	\monad \monad D 				&	\rTo_{\id{\monad\monad D}} 	& \monad \monad D						
\end{diagram}

%% file: modules/diagrams/GeneralFramework/MonadTriangularIdentityNeutral.tex
%\hfill\\
\begin{diagram}[LaTeXeqno,righteqno,balanced,heads=littlevee]
\label{diagrams_GeneralFramework_MonadTriangularIdentityNeutral}
\monad \monad A				&							&		\\
\uEmbed<{\unit{\monad A}}	&	\rdOnto^{\mult{A}}		&		\\
\monad A					& 	\rTo_{\id{\monad A}}	&	\monad A 					
\end{diagram}
%\hfill\\

%% file: modules/diagrams/GeneralFramework/MonadSquareIdentity.tex
\begin{diagram}[LaTeXeqno,righteqno,balanced,heads=littlevee]
\label{diagrams_GeneralFramework_MonadSquareIdentity}
\monad \monad \monad A		&	\rOnto^{\monad \mult{A}}	&	\monad \monad A		\\
\dOnto<{\mult{\monad A}} 	&								&	\dOnto>{\mult{A}}		\\
\monad \monad A				&	\rOnto_{\mult{A}}			&	\monad A 			
\end{diagram}

%% file: modules/diagrams/GeneralFramework/MonadTriangularIdentityInitSurfExplained.tex
\begin{diagram}[LaTeXeqno,righteqno,balanced,heads=littlevee]
\label{diagrams_GeneralFramework_MonadTriangularIdentityInitSurf}
	&							&\monad D 				&\rTo{\freehistory{\unit{A} \circ d}}	&\monad \monad A				&						&				\\	
	&\ruDashembed(2,4)			&\uTo>{\id{\monad D}}	&\ruDashembed(1,4)						&\uEmbed>{\monad \unit{ A}}		&\rdOnto^{\mult{A}}		&				\\
	&							&\monad D 				&\rTo^{\freehistory{d}} 				&\monad A						&\rTo_{\id{\monad A}}	&\monad A 		\\	
	&\ruDashembed(2,4)			& 						&\ruDashembed(1,4)						&								&						&\dOnto>{\alpha}\\	
D				&				&\rTo(1,0){\unit{A} \circ d}	&\monad A						&								&						&A				\\	
\uTo<{\id{D}}	&				&								&\uEmbed<{\unit{A}}				&								&						&				\\
D				&				&\rTo(1,0){d}					&A								&								&						&				
\end{diagram}

%% file: modules/diagrams/GeneralFramework/MonadTriangularIdentityInitSurf.tex
\begin{diagram}[LaTeXeqno,righteqno,balanced,heads=littlevee]
\label{diagrams_GeneralFramework_MonadTriangularIdentityInitSurf}
\monad \monad A			&							&						\\
\uEmbed<{\monad \unit{ A}}	&	\rdOnto^{\mult{A}}		&						\\
\monad A				&	\rTo_{\id{A}}			&	\monad A 			\\
\end{diagram}

%% file: modules/diagrams/GeneralFramework/EMCategoryDynSymmetries.tex
\begin{diagram}[LaTeXeqno,righteqno,balanced,heads=littlevee]
\label{diagrams_GeneralFramework_AlgebraFreeDynamicalStructure}
\monad  A 		&	\rTo_{\monad \phi}^\iso  	&	\monad A 				\\
\dOnto<{\alpha}	&								&	\dOnto>{\alpha}			\\
 A 				&	\rTo_{\phi}^\iso			&	A 						
\end{diagram}

%% file: modules/pictures/TimeActionAssociativityTikZ.tex
\begin{tikzpicture}

\draw (-1,-0.5) -- (-1,4.5) -- (8,4.5) -- (8,-0.5) -- (-1,-0.5);

%% Layer -1 (inputs)
	\node[boundary]	(input_timeLHSa) 	at (0,0) 	{};
	\node[boundary]	(input_timeLHSb) 	at (2,0)	{};
	\node[boundary]	(input_timeLHSc) 	at (3,0) 	{};

	\node[boundary]	(input_timeRHSa) 	at (4,0) 	{};
	\node[boundary]	(input_timeRHSb) 	at (5,0)	{};
	\node[boundary]	(input_timeRHSc) 	at (7,0) 	{};

%% Layer 0
	\node[spider] 	(timeActionLHS0) 	at (1,1.5)	{$\nu$};
	\node[spider] 	(timeActionRHS0) 	at (6,1.5)	{$\nu$};

%% Equality sign
	\node			(equality)			at (3.5,2.5){$=$};

%% Layer 1
	\node[spider] (timeActionLHS1) 	at (2,3)	{$\nu$};
	\node[spider] (timeActionRHS1) 	at (5,3)	{$\nu$};

%% Layer 2 (outputs)
	\node[boundary]		(output_timeLHS)	at (2,4)	{};
	\node[boundary]		(output_timeRHS)	at (5,4)	{};

%% Transitions to 0 	
	\draw[system] 	(input_timeLHSa)	to [out=90,in=225] 	(timeActionLHS0);
	\draw[system] 	(input_timeLHSb)	to [out=90,in=315] 	(timeActionLHS0);

	\draw[system] 	(input_timeRHSb)	to [out=90,in=225] 	(timeActionRHS0);
	\draw[system]	(input_timeRHSc)	to [out=90,in=315] 	(timeActionRHS0);

%% Transitions to 1
	\draw[system] 	(timeActionLHS0)	to [out=90,in=225] 	(timeActionLHS1);
	\draw[system] 	(input_timeLHSc)	to [out=90,in=315] 	(timeActionLHS1);

	\draw[system] 	(timeActionRHS0)	to [out=90,in=315] 	(timeActionRHS1);
	\draw[system] 	(input_timeRHSa)	to [out=90,in=225] 	(timeActionRHS1);

%% Transitions to 2 (outputs)
	\draw[system] 	(timeActionLHS1) 	to [out=90,in=270]	(output_timeLHS);

	\draw[system] 	(timeActionRHS1) 	to [out=90,in=270]	(output_timeRHS);

\end{tikzpicture}

%% file: modules/pictures/TimeActionUnitTikZ.tex
\begin{tikzpicture}

\draw (-1,-0.5) -- (-1,4.5) -- (5,4.5) -- (5,-0.5) -- (-1,-0.5);

%% Layer -1: inputs
	\node[boundary]	(input_time) 	at (0,0) {};
	\node[boundary] (input_trivial)	at (2,0)	{};
	\node[boundary]	(input_timeRHS) at (3,0)	{};

%% Layer 0
	\node[spider] 	(timeUnit)		at (2,1.5)	{$\eta_I$};

%% Layer 1
	\node[spider] 		(timeAction) 	at (1,3)	{$\nu$};
	\node 				(equals) 		at (2.5,2.5) 	{$=$};

%% Layer 2: outputs					
	\node[boundary]		(output_time)	 at (1,4)	{};
	\node[boundary] 	(output_timeRHS) at (4,4)   {};

%% Transitions from -1 to 0 	
	\draw[system] 		(input_time)	to [out=90,in=225] 	(timeAction);
	\draw[trivial] 	(input_trivial)	to [out=90,in=270]	(timeUnit);
	\draw[system] 		(input_timeRHS)	to [out=90,in=270] 	(output_timeRHS);

%% Transitions from 0 to 1
	\draw[system] 		(timeUnit) 		to [out=90,in=315]	(timeAction);

%% Transitions from 1 to 2
	\draw[system] 		(timeAction) 	to [out=90,in=270]	(output_time);

\end{tikzpicture}

%% file: modules/diagrams/SymmetricMonoidalFramework/TimeActionAssociativity.tex
\begin{diagram}[LaTeXeqno,righteqno,balanced,heads=littlevee]
\label{diagrams_SymmetricMonoidalFramework_TimeActionAssociativity}
\timeobj \tensor \timeobj \tensor \timeobj 	& \rTo^{\id{\timeobj}\tensor\foliate{\timeobj}}	& \timeobj \tensor \monad \timeobj	& \rTo^{\id{\timeobj}\tensor\mult{\tensorUnit}}	& \timeobj \tensor \timeobj \\
\dTo<{\foliate{\timeobj}\tensor\id{\timeobj}}& \text{see below}								& \dTo								& t \text{ respects } \mult{}					& \dTo>{\foliate{\timeobj}} \\
\monad \timeobj \tensor \timeobj			& \rTo^{\foliate{\monad \timeobj}}				& \monad \monad \timeobj			& \rTo^{\mult{\monad \tensorUnit}}				& \monad \timeobj 			\\
\dTo<{\mult{\tensorUnit} \tensor \id{\timeobj}}	& \begin{array}{c} \text{naturality of } \\ \text{foliation} \end{array} & \dTo~{\monad \mult{\tensorUnit}}  & \begin{array}{c} \text{associat.} \\ \text{of } \mult{} \end{array} & \dTo>{\mult{\tensorUnit}}	\\
\timeobj \tensor \timeobj					& \rTo_{\foliate{\timeobj}}						& \monad \timeobj					& \rTo_{\mult{\tensorUnit}}						& \timeobj					
\end{diagram}

%% file: modules/diagrams/SymmetricMonoidalFramework/TimeActionAssociativityDetail.tex
\begin{diagram}[LaTeXeqno,righteqno,balanced,heads=littlevee]
\label{diagrams_SymmetricMonoidalFramework_TimeActionAssociativityDetail}
\timeobj \tensor \timeobj \tensor \timeobj 				& \rTo^{\id{\timeobj}\tensor\foliate{\timeobj}}		& \timeobj \tensor \monad \timeobj	\\
\dTo<{\id{\timeobj \tensor \timeobj \tensor \timeobj}}	& \text{associat. of } t							& \dTo>{t_{\timeobj,\timeobj}}		\\
\timeobj \tensor \timeobj \tensor \timeobj 				& \rTo^{\foliate{\timeobj \tensor \timeobj}}		& \monad (\timeobj \tensor \timeobj)\\
\dTo<{\foliate{\timeobj} \tensor \id{\timeobj}}			& \text{naturality of } \text{foliation}					& \dTo>{\monad \foliate{\timeobj}}  \\
\monad \timeobj \tensor \timeobj						& \rTo_{\foliate{\monad \timeobj}}					& \monad \monad \timeobj			\\
\end{diagram}

%% file: modules/diagrams/SymmetricMonoidalFramework/FoliationRespectsUnit.tex
\begin{diagram}[LaTeXeqno,righteqno,balanced,heads=littlevee]
\label{diagrams_SymmetricMonoidalFramework_FoliationRespectsUnit}
A \tensor \timeobj 							&	\rTo^{\foliate{A}} 	&	\monad A 				\\
\uEmbed<{\id{A} \tensor \bar{\timeunit}}	&						&	\uEmbed>{\unit{A}}		\\
A											&	\rTo_{\id{A}} 		&	A 						
\end{diagram}

%% file: modules/diagrams/SymmetricMonoidalFramework/FoliationRespectsMult.tex
\begin{diagram}[LaTeXeqno,righteqno,balanced,heads=littlevee]
\label{diagrams_SymmetricMonoidalFramework_FoliationRespectsMult}
A \tensor \timeobj \tensor \timeobj		&	\rTo^{\monad \foliate{A}\cdot \foliate{A \tensor \timeobj}}_{\foliate{\monad A} \cdot (\foliate{A} \tensor \id{\timeobj})} 	&	\monad \monad A 		\\
\dOnto<{\id{A} \tensor\bar{\timemult}}					&																		&	\dOnto>{\mult{A}}		\\
A \tensor \timeobj						&	\rTo_{\foliate{A}} 													&	\monad A 						
\end{diagram}

%% file: modules/diagrams/SymmetricMonoidalFramework/LiftedAlgebrasIVP.tex
\begin{diagram}[LaTeXeqno,righteqno,balanced,heads=littlevee]
\label{diagrams_SymmetricMonoidalFramework_LiftedAlgebrasIVP}
A \tensor \timeobj 							&	\rTo^{\foliate{A}} 	&	\monad A 				\\
\uEmbed<{\id{A} \tensor \unit{\tensorUnit}}	&	\ruEmbed_{\unit{A}}	&	\dOnto>{\alpha}			\\
A											&	\rTo_{\id{A}} 		&	A 						
\end{diagram}

%% file: modules/diagrams/SymmetricMonoidalFramework/LiftedAlgebrasFreeDynamics.tex
\begin{diagram}[LaTeXeqno,righteqno,balanced,heads=littlevee]
\label{diagrams_SymmetricMonoidalFramework_LiftedAlgebrasFreeDynamics}
A \tensor \timeobj \tensor \timeobj 	&&& \rTo(3,0)^{\foliate{A}\tensor\id{\timeobj}}	& \monad A \tensor \timeobj	& \rTo^{\alpha \tensor \id{\timeobj}}	& A \tensor \timeobj\\
\dTo<{\id{A} \tensor \foliate{\timeobj}}&& \text{see below}							&& \dTo~{\foliate{\monad A}}	& \begin{array}{c} \text{ naturality} \\ \text{of foliation} \end{array}		& \dTo>{\foliate{A}}\\
A \tensor \monad \timeobj				&\rTo^{t_{A,\timeobj}}& \monad (A \tensor \timeobj)& \rTo^{\monad \foliate{A}}									& \monad \monad A			& \rTo^{\monad \alpha}					& \monad A 			\\
\dTo<{\id{A} \tensor \mult{\tensorUnit}}&& t \text{ respects }\mult{} 				&& \dTo~{\mult{A}}  			& \text{associat. of }\alpha 			& \dTo>{\alpha}		\\
A \tensor \timeobj						&&& \rTo_{\foliate{A}}(3,0)						& \monad A					& \rTo_{\alpha}							& \timeobj				
\end{diagram}